\documentclass[a4paper,twocolumn,11pt,accepted=2022-05-13]{quantumarticle}
\pdfoutput=1
\usepackage[utf8]{inputenc}
\usepackage[english]{babel}
\usepackage[T1]{fontenc}
\usepackage{hyperref}

\usepackage{amsmath,amssymb,amsthm}
\usepackage{braket}
\usepackage{cleveref}
\usepackage{graphicx}
\usepackage[caption=false]{subfig}
\usepackage{float}
\usepackage{tabularx}
\usepackage{amsfonts}
\usepackage{xcolor}

\usepackage{tikz}
\usepackage{lipsum}

\newtheorem{definition}{Definition}

\begin{document}

\title{Quantum agents in the Gym:

a variational quantum algorithm for deep Q-learning}

\author{Andrea Skolik}
    \affiliation{Leiden University, Niels Bohrweg 1, 2333 CA Leiden, The Netherlands}
    \affiliation{Volkswagen Data:Lab, Ungererstra\ss e 69, 80805 Munich, Germany}

\author{Sofiene Jerbi}
    \affiliation{Institute for Theoretical Physics, University of Innsbruck, Technikerstr. 21a, A-6020 Innsbruck, Austria}
    
\author{Vedran Dunjko}
    \affiliation{Leiden University, Niels Bohrweg 1, 2333 CA Leiden, The Netherlands}

\begin{abstract}
  Quantum machine learning (QML) has been identified as one of the key fields that could reap advantages from near-term quantum devices, next to optimization and quantum chemistry. Research in this area has focused primarily on variational quantum algorithms (VQAs), and several proposals to enhance supervised, unsupervised and reinforcement learning (RL) algorithms with VQAs have been put forward. Out of the three, RL is the least studied and it is still an open question whether VQAs can be competitive with state-of-the-art classical algorithms based on neural networks (NNs) even on simple benchmark tasks. In this work, we introduce a training method for parametrized quantum circuits (PQCs) that can be used to solve RL tasks for discrete and continuous state spaces based on the deep Q-learning algorithm.
  We investigate which architectural choices for quantum Q-learning agents are most important for successfully solving certain types of environments by performing ablation studies for a number of different data encoding and readout strategies. We provide insight into why the performance of a VQA-based Q-learning algorithm crucially depends on the observables of the quantum model and show how to choose suitable observables based on the learning task at hand. To compare our model against the classical DQN algorithm, we perform an extensive hyperparameter search of PQCs and NNs with varying numbers of parameters. We confirm that similar to results in classical literature, the architectural choices and hyperparameters contribute more to the agents' success in a RL setting than the number of parameters used in the model. Finally, we show when recent separation results between classical and quantum agents for policy gradient RL can be extended to inferring optimal Q-values in restricted families of environments. This work paves the way towards new ideas on how a quantum advantage may be obtained for real-world problems in the future.
\end{abstract}

\section{Introduction}

Variational quantum algorithms are among the most promising candidates to show quantum advantage in the near-term. Quantum machine learning has emerged as one field that is amenable to applications of VQAs on noisy intermediate-scale quantum (NISQ) devices \cite{bharti2021noisy,preskill2018quantum}. Many proposals for QML algorithms have been made in supervised \cite{mitarai2018quantum,schuld2020circuit,schuld2019quantum,havlivcek2019supervised,farhi2018classification} and unsupervised \cite{amin2018quantum,coyle2020born,zoufal2020variational,lloyd2018quantum,zoufal2019quantum,chakrabarti2019quantum} learning. In contrast, RL is a subfield of machine learning that has received less attention in the QML community \cite{hamann2020quantum,jerbi2019quantum}, and especially proposals for VQA-based approaches are only now emerging \cite{chen2020variational,lockwood2020reinforcement,wu2020quantum,benedetti2019parameterized,jerbi2021variational}. RL is essentially a way to solve the problem of optimal control. In a RL task, an agent is not given a fixed set of training data, but learns from interaction with an environment. Environments are defined by a space of states they can be in, and a space of actions that an agent uses to alter the environment's state. The agent chooses its next action based on a policy (probability distribution over actions given states) and receives a reward at each step, and the goal is to learn an optimal policy that maximizes the long-term reward the agent gets in the environment. State and action spaces can be arbitrarily complex, and it's an open question which types of models are best suited for these learning tasks. In classical RL, using NNs as function approximators for the agents' policy has received increased interest in the past decade. As opposed to learning exact functions to model agent behavior which is infeasible in large state and action spaces, this method of RL only approximates the optimal function. These types of RL algorithms have been shown to play Atari arcade games as well as human players \cite{mnih2015human}, and even reach super-human levels of performance on games as complex as Go \cite{silver2016mastering}, Dota \cite{berner2019dota} and StarCraft \cite{vinyals2019grandmaster}. RL algorithms can be divided into \textit{policy-based} and \textit{value-based} methods. While policy-based algorithms seek to directly optimize a parametrized probability distribution that represents the policy, a value-based algorithm learns a so-called value function which then gives rise to a policy. These two methods constitute related but fundamentally different approaches to solve RL tasks, and both have their own (dis-)advantages as we will explain in detail in \cref{sec:value_vs_policy}. Interestingly, these two methods can also be combined in a so-called \textit{actor-critic} setting which leverages the strengths of both approaches \cite{konda2000actor}. Actor-critic methods are among the state-of-the-art in current RL literature \cite{mnih2016asynchronous}, and therefore both value-based and policy-based algorithms are areas of active research. 

One classical value-based RL algorithm that has gained much popularity is called \textit{Q-learning} \cite{watkins1989learning}, where the goal is to learn \textit{Q-values}, numerical values which for a given state-action pair represent the expected future reward. Based on these values, a Q-learning agent chooses the appropriate action in a given state, where a higher Q-value corresponds to a higher expected reward. The NN's role in a Q-learning algorithm is to serve as a function approximator for the Q-function. It is thus natural to ask whether RL algorithms can be adapted to work with a VQA approach, and whether such algorithms could offer an advantage over their classical counterparts. RL is one of the hardest modes of learning in current ML research, and is known to require careful tuning of model architectures and hyperparameters to perform well. For NN-based approaches, one unfavorable hyperparameter setting can lead to complete failure of the learning algorithm on a specific task. Additionally, these hyperparameters and architectures are highly task dependent and there is no a-priori way to know which settings are best. Well-performing settings are found by experts via trial-and-error, and the ability to quickly find these settings is considered a ``black art that requires years of experience to acquire'' \cite{smith2018disciplined}. Thus a whole field of heuristics and numerical studies has formed on finding good sets of hyperparameters like NN architectures \cite{ye2019comparison,yu2011comparison,cordoni2020comparison}, activation functions \cite{szandala2021review,nwankpa2018activation,urban2018neural}, or learning rates and batch sizes \cite{smith2018disciplined,smith2017cyclical}. An increasingly investigated branch of research focuses on methods to automate the whole process of finding good architectures and hyperparameters, among which there is neural architecture search \cite{elsken2019neural} and automated machine learning \cite{hutter2019automated}. 

It is thus to be expected that quantum models in a VQA-based RL setting also need to be selected carefully. Even more so, it is still an open question whether VQAs are suitable for function approximation in RL at all. This question is directly related to choices made when defining an architecture for a VQA. There are three important factors to consider: the structure (or \textit{ansatz}) of the model, the data-encoding technique, and the readout operators. For the choice of structure, there is a trade-off between the expressivity and trainability of a model, as certain structures are subject to the so-called barren plateau phenomenon \cite{mcclean2018barren}. This phenomenon prevents successful training of models with a large number of qubits and layers for highly expressive structures like random circuits. On the other hand, overparametrization has been observed to simplify optimization landscapes and lead to faster convergence for certain VQAs \cite{kiani2020learning,wiersema2020exploring}. Apart from that, the choice of structure is also limited by hardware constraints like the topology of a certain quantum device. While the model structure is an important factor in training VQAs that has received much attention in the QML community \cite{cerezo2020cost,wang2020noise,skolik2020layerwise,marrero2020entanglement,sim2019expressibility,sim2021adaptive,liu2021layer}, the authors of \cite{schuld2020effect} have shown that the technique used to encode data into the model plays an equally important role, and that even highly expressive structures fail to fit simple functions with an insufficient data-encoding strategy. 

A less explored architectural choice in the context of QML is that of the observables used to read out information from the quantum model. Considering that the readout operator of a quantum model fixes the range of values it can produce, this choice is especially important for tasks where the goal is to fit a real-valued function with a given range, as is the case in many RL algorithms. This is in contrast to NNs, which have no restriction on the range of output values and can even change this range dynamically during training. In Q-learning, the goal is to approximate the real-valued optimal Q-function, which can have an arbitrary range based on the environment. Crucially, this range can change depending on the performance of the agent in the environment, which is an impediment for quantum models with a fixed range of output values.

A first step to study the influence of architectural choices on PQCs for policy-based RL algorithms has been made in \cite{jerbi2021variational}, who point out that data-encoding and readout strategies play a crucial role in these types of RL tasks, though they leave the open question if similar architectural choices are also required in a value-based setting. Previous work on Q-learning with PQCs has addressed certain other fundamental questions about the applicability of VQAs in a value-based context. A VQA for Q-learning in discrete state spaces was introduced in \cite{chen2020variational}, where the quantum model's output is followed by a layer of additive weights, and it has been shown that the model successfully solves two discrete-state environments. A VQA for Q-learning in environments with continuous and discrete state spaces has been proposed in \cite{lockwood2020reinforcement}, who simplify the continuous environments' potentially infinite range of input values to a restricted encoding into angles of one initial layer of rotation gates, and use measurements in the Z-basis to represent Q-values. Notably, none of the models in \cite{lockwood2020reinforcement} that were run for the continuous state-space environment Cart Pole reach a performance that is considered to be solving the environment according to its original specification \cite{cp_wiki}, so it remains an open question whether a value-based algorithm that utilizes a PQC as the function approximator can solve this type of learning task. 

These initial works prompt a number of vital follow-up questions related to the architectural choices that are required to succeed in arbitrary RL environments with a quantum Q-learning agent. We address these questions in form of our main contributions as follows: first, we propose a VQA which can encode states of discrete and continuous RL environments and explain the intricate relationship between the environment's specification and the requirements on the readout operators of the quantum model. We show how a quantum Q-learning agent only succeeds if these requirements are met. Second, to enable the model to match the environment's requirements on the range of output values, we make this range itself trainable by introducing additional weights on the model outputs. We show how the necessity of these weights can be inferred from the range that the optimal Q-values take in an environment. Third, we study the performance of our model on two benchmark environments from the OpenAI Gym \cite{brockman2016openai}, Frozen Lake and Cart Pole. For the continuous-state Cart Pole environment, we also study a number of data encoding methods and illustrate the benefit of previously introduced techniques to increase quantum model expressivity, like data re-uploading \cite{perez2020data} or trainable weights on the input data \cite{perez2020data,jerbi2021variational}. Additionally, the state space dimension of both environments is small enough so that inputs can be directly encoded into the quantum model without the use of a dimensionality reduction technique. This makes it possible to directly compare our model to a NN performing the same type of Q-learning algorithm to evaluate its performance. Specifically, we perform an in-depth comparison of the performance of PQCs and NNs with varying numbers of parameters on the Cart Pole environment. We show that recent results in classical deep Q-learning also apply to the case when a PQC is used as the function approximator, namely that increasing the number of parameters is only beneficial up to some point \cite{ota2021training}. After this, learning becomes increasingly unstable for both PQCs and NNs. As an empirical comparison between PQCs and NNs can only give us insight into model performance on the specific environments we study, we also explain when recent separation results for policy gradient RL between classical and quantum agents \cite{jerbi2021variational} also hold in the Q-learning setting for restricted families of environments.

The remainder of this paper is structured as follows: in \cref{sec:rl} we give an introduction to RL, followed by a description of our quantum RL model in \cref{sec:qrl}. We show when recent results for a separation between classical and quantum algorithms for policy-based learning also apply in the case of Q-learning in \cref{sec:separation}. In \cref{sec:numerics} we numerically evaluate the performance of our algorithm and compare it to a classical approach, and finally discuss our findings in \cref{sec:conclusion}. The full code that was used to perform the numerical experiments in this work can be found on Github \cite{quantum_agents_code}.

\section{Reinforcement learning}
\label{sec:rl}

In RL, an agent does not learn from a fixed data set as in other types of learning, but by making observations on and interacting with an environment \cite{sutton2018reinforcement}. This distinguishes it from the other two main branches of ML, supervised and unsupervised learning, and each of the three comes with its individual challenges. In a supervised setting, an agent is given a fixed set of training data that is provided with the correct labels, where difficulties arise mainly in creating models that do not overfit the training data and keep their performance high on unseen samples. In unsupervised learning, training data is not labeled and the model needs to discover the underlying structure of a given data set, and the challenge lies in finding suitable loss functions and training methods that enable this. RL also comes with a number of challenges: there is no fixed set of training data, but the agent generates its own samples by interacting with an environment. These samples are not labeled, but only come with feedback in form of a reward. Additionally, the training data keeps changing throughout the learning process, as the agent constantly receives feedback from interacting with its environment.

An environment consists of a set of possible states $\mathcal{S}$ that it can take, and a set of actions $\mathcal{A}$ which the agent can perform to alter the environment's state. Both state and action spaces can be continuous or discrete. An agent interacts with an environment by performing an action $a_t$ at time step $t$ in state $s_t$, upon which it receives a reward $r_{t+1}$. A tuple $(s, a, r, s')$ of these four quantities is called a \textit{transition}, and transition probabilities from state $s$ to $s'$ after performing action $a$ in a given environment are represented by the \textit{transition function} $P_{ss'}^a$,
\begin{equation}
    P_{ss'}^a = P(s'|s, a).
\end{equation}
The reward function is designed to evaluate the quality of the agent's actions on the environment based on the learning task at hand, and the agent's goal is to maximize its total reward over a sequence of time steps starting at $t$, called the return $G_t$
\begin{equation}\label{eq:return}
    G_t = \sum_{k=0}^{\infty} \gamma^k r_{t+k+1},
\end{equation}
where $\gamma \in [0, 1]$ is a discount factor introduced to prevent divergence of the infinite sum. The return $G_t$ should be viewed as the agent's expected reward when starting from time step $t$ and summing the discounted rewards of potentially infinitely many future time steps, where maximizing the return at step $t$ implies also maximizing the return of future time steps. Note that the task is to maximize an expected value, and that the reward $r_t$ in \cref{eq:return}, and therefore $G_t$ are random variables. Environments often naturally break down into so-called \textit{episodes}, where the sum in \cref{eq:return} is not infinite, but only runs over a fixed number of steps called \textit{horizon} $H$. An example of this are environments based on games, where one episode comprises one game played and an agent learns by playing a number of games in series.

\subsection{Value-based and policy-based learning}
\label{sec:value_vs_policy}

RL algorithms can be categorized into \textit{value-based} and \textit{policy-based} learning methods \cite{sutton2018reinforcement}. Both approaches aim to maximize the return as explained above, but use different figures of merit to achieve this. Both approaches also have their disadvantages as we will see below, and which type of algorithm should be used depends on the environment at hand. In both cases, the function that models the agent's behavior in the environment is called the \textit{policy} $\pi(a|s)$, which gives the probability of taking action $a$ in a given state $s$. The main difference between the two approaches is how the policy is realized. In general, performance is evaluated based on a state-value function (or an action-value function, as we will see in \cref{sec:q-learning}) $V_{\pi}(s)$,
\begin{equation}\label{eq:val_function}
    V_{\pi}(s) = \mathbb{E}_{\pi}[G_t|s_t = s],
\end{equation}
which is the \textit{expected return} when following policy $\pi$ starting from state $s$ at initial time step $t$, and the goal of a RL algorithm is to learn the optimal policy $\pi_*$ which maximizes the expected return for each state.

A policy-based algorithm seeks to learn an optimal policy directly, that is, learn a probability distribution of actions given states. In this setting, the policy is implemented in form of a parametrized conditional probability distribution $\pi(a|s;\boldsymbol{\theta})$, and the goal of the algorithm is to find parameters $\boldsymbol{\theta}$ such that the resulting policy is optimal. The figure of merit in this setting is some performance measure $J(\boldsymbol{\theta})$ that we seek to maximize. This performance measure can include a value function as in \cref{eq:val_function}, however, once the policy has been learned $J(\boldsymbol{\theta})$ is not required for action generation. Typically, these algorithms perform gradient ascent on an approximation of the gradient of the performance measure $\nabla J(\boldsymbol{\theta})$, which is obtained by Monte Carlo samples of policy rollouts (i.e., a set of observed interactions with the environment performed under the given policy), and are hence called \textit{policy gradient} methods. This approach produces smooth updates on the policy (as opposed to value-based algorithms, where a small change in the value function can drastically alter the policy) that enable proofs of convergence to locally optimal policies \cite{sutton1999policy}. However, it also suffers from high variance as updates are purely based on Monte Carlo samples of interactions with the environment \cite{greensmith2004variance}. A number of methods to reduce this variance have been developed, like adding a value-based component as described below to a policy-based learner in the so-called \textit{actor-critic} method \cite{konda2000actor}.

In a value-based algorithm, a value function as in \cref{eq:val_function} is learned instead of the policy. The policy is then implicitly given by the value function: an agent will pick the action which yields the highest expected return according to $V_{\pi}(s)$. A concrete example of value-based learning is given in \cref{sec:q-learning}, where we introduce the algorithm that we focus on in this work. While value-based algorithms do not suffer from high training variance as policy gradient learning does, they often require more episodes to converge. They also result in deterministic policies, as the agent always picks the action that corresponds to the highest expected reward, so this approach will fail when the optimal policy is stochastic and post-training action selection is performed according to the argmax policy.\footnote{Consider for example a game of poker where bluffing is a valid action to scare other players into folding, but quickly becomes obvious when greedily done in every round.} Additionally, the policy resulting from a parametrized value function can change substantially after a single parameter update (i.e., a very small change in the value function can lead to picking a different action after an update). This results in theoretical difficulties to prove convergence when a function approximator is used to parametrize the value function, hence there are even fewer theoretical guarantees for this approach than for policy gradient methods. On the other hand, it was the advent of deep Q-learning that made it possible to solve extremely complex problems such as Go with a reinforcement learning approach \cite{silver2016mastering}.

Both approaches have their own (dis-) advantages, and while the popularity of either method has surpassed the other at some point in the last decades, there is no clear winner. As mentioned above, an actor-critic approach combines a policy-based and value-based learner to leverage the advantages of both while alleviating the disadvantages, and this method is among the stat-of-the-art in classical RL literature \cite{mnih2016asynchronous}. Additionally, it can be easier to learn either the policy or the value function depending on a given environment. For this reason, both approaches are worth being studied independently. In the quantum setting, VQAs for policy gradient learning have been investigated in \cite{wu2020quantum,jerbi2021variational}. Using PQCs for value-based learning has been explored in \cite{chen2020variational,lockwood2020reinforcement}, and a specific value-based RL algorithm called \textit{Q-learning} is the focus of this work.

\subsection{Q-learning}
\label{sec:q-learning}

In Q-learning, we are not interested in the state-value function as shown in \cref{eq:val_function}, but in the closely related action-value function $Q_{\pi}(s, a)$,
\begin{equation}\label{eq:q_exp_return}
    Q_{\pi}(s, a) = \mathbb{E}_{\pi}[G_t|s_t = s, a_t = a],
\end{equation}
which also gives us the expected return assuming we follow a policy $\pi$, but now additionally conditioned on an action $a$. We call the optimal Q-function $Q_*(s, a) = \max_{\pi} Q_{\pi}(s, a)$, and an optimal policy can be easily derived from the optimal values by taking the highest-valued action in each step, as
\begin{equation}
    \pi_*(a|s) = \underset{a}{\mathrm{argmax}}~Q_*(s, a).
\end{equation}
The goal in Q-learning is to learn an estimate, $Q(s, a)$, of the optimal Q-function. In its original form, Q-learning is a tabular learning algorithm, where a so-called \textit{Q-table} stores Q-values for each possible state-action pair \cite{watkins1989learning}. When interacting with an environment, an agent chooses its next action depending on the Q-values as
\begin{equation}\label{eq:q_action_choice}
    a_t= \underset{a}{\mathrm{argmax}}~Q(s_t, a),
\end{equation}
where a higher value designates a higher expected reward when action $a$ is taken in state $s_t$ as opposed to the other available actions. When we consider learning by interaction with an environment, it is important that the agent is exposed to a variety of transitions to sufficiently explore the state and action space. Intuitively, this provides the agent with enough information to tell apart good and bad actions given certain states. Theoretically, visiting all state-action transitions infinitely often is one of the conditions that are required to hold for convergence proofs of tabular Q-learning to an optimal policy \cite{melo2001convergence}. Clearly, if we always follow an argmax policy, the agent only gets access to a limited part of the state and action space. To ensure sufficient exploration in a Q-learning setting, a so-called $\epsilon$\textit{-greedy policy} is used. That is, with probability $\epsilon$ a random action is performed and with probability $1 - \epsilon$ the agent chooses the action which corresponds to the highest Q-value for the given state as in \cref{eq:q_action_choice}. Note that the $\epsilon$-greedy policy is only used to introduce randomness to the actions picked by the agent during training, but once training is finished, a deterministic argmax policy is followed.

The Q-values are updated with observations made in the environment by the following update rule, 

\begin{equation}\label{eq:q_learning_update}
\begin{split}
    Q(s_t, a_t) \leftarrow Q(s_t, a_t) + \alpha[r_{t+1} + \gamma \cdot \underset{a}{\max}~Q(s_{t+1}, a)\\
    - Q(s_t, a_t)],
\end{split}
\end{equation}
where $\alpha$ is a learning rate, $r_{t+1}$ is the reward at time $t+1$, and $\gamma$ is a discount factor. Intuitively, this update rule provides direct feedback from the environment in form of the observed reward, while simultaneously incorporating the agent's own expectation of future rewards at the present time step via the maximum achievable expected return in state $s_{t+1}$. In the limit of visiting all $(s, a)$ pairs infinitely often, this update rule is proven to converge to the optimal Q-values in the tabular case \cite{melo2001convergence}. 

Q-values can take an arbitrary range, which is determined by the environment's reward function and the discount factor $\gamma$, which controls how strongly expected future rewards influence the agent's decisions. Depending on $\gamma$, the optimal Q-values for the same environment can take highly varying values, and can therefore be viewed as different learning environments themselves. In practice, it is not necessary that an agent learns the optimal Q-values exactly. As the next action at step $t$ is chosen according to \cref{eq:q_action_choice}, it is sufficient that the action with the highest expected reward has the highest Q-value for the sake of solving an environment presuming a deterministic policy. In other words, for solving an environment only the $order$ of Q-values is important, and the task is to learn this correct order by observing rewards from the environment through interaction. As our goal in this work is to solve certain environments with Q-learning, learning the correct order of Q-values is sufficient and we restrict our attention to this case throughout our numerical studies.

Obviously, the tabular approach is intractable for large state and action spaces. For this reason, the Q-table was replaced in subsequent work by a Q-function approximator which does not store all Q-values individually \cite{lin1992self,melo2007q}. In the seminal work \cite{mnih2015human}, the authors use a NN as the Q-function approximator which they call \textit{deep Q-network} (DQN) and the resulting algorithm the \textit{DQN algorithm}, and demonstrate that this algorithm achieves human-level performance on a number of arcade games. In this work, the agent chooses actions based on an $\epsilon$-greedy policy as described above. Typically $\epsilon$ is chosen large in the beginning and then decayed in subsequent iterations, to ensure that the agent can sufficiently explore the environment at early stages of training by being exposed to a variety of states. The authors of \cite{mnih2015human} also utilize two other improvements over previous approaches which we adopt: (i) \textit{experience replay}: past transitions and their outcomes are stored in a memory, and the batches of these transitions that are used to compute parameter updates are sampled at random from this memory to remove temporal correlations between transitions, (ii) adding a second so-called \textit{target network} to compute the expected Q-values for the update rule, where the target network has an identical structure as the DQN, but is not trained and only sporadically updated with a copy of the parameters of the DQN to increase stability during training. We refer to the original paper for further details on the necessity of these techniques to stabilize training of the DQN algorithm. 

The DQN is then trained almost in a supervised fashion, where the training data and labels are generated by the DQN itself through interaction with the environment. At each update step, a batch $\mathcal{B}$ of previous transitions $(s_t, a_t, r_{t+1}, s_{t+1})$ is chosen from the replay memory. To perform a model update, we need to compute $\underset{a}{\max}~Q(s_{t+1}, a)$. When we use a target network, this value is not computed by the DQN, but by the target network $\hat{Q}$. To make training more efficient, in practice the Q-function approximator is redefined as a function of a state parametrized by $\boldsymbol{\theta}$, $Q_{\boldsymbol{\theta}}(s) = \boldsymbol{q}$, which returns a vector of Q-values for all possible actions instead of computing each $Q(s, a)$ individually. We now want to perform a supervised update of $Q_{\boldsymbol{\theta}}$, where the label is obtained by applying the update rule in \cref{eq:q_update_dqn} to the DQN's output. To compute the label for a state $s$ that we have taken action $a$ on in the past, we take a copy of $Q_{\boldsymbol{\theta}}(s)$ which we call $\boldsymbol{q_{\delta}}$, and only the $i$th entry of $\boldsymbol{q_{\delta}}$ is altered where $i$ corresponds to the index of the action $a$, and all other values remain unchanged. The estimated maximum Q-value for the following state $s_{t+1}$ is computed by $\hat{Q}_{\boldsymbol{\theta}_{\delta}}$, and the update rule for the $i$-th entry in $\boldsymbol{q_{\delta}}$ takes the following form
\begin{equation}\label{eq:q_update_dqn}
     \boldsymbol{q}_{\delta_i} = r_{t+1} + \gamma \cdot  \underset{a}{\max}~\hat{Q}_{\boldsymbol{\theta}_{\delta}}(s_{t+1}, a),
\end{equation}
where $\boldsymbol{\theta}_{\delta}$ is a periodically updated copy of $\boldsymbol{\theta}$.
The loss function $\mathcal{L}$ is the mean squared error (MSE) between $\boldsymbol{q}$ and $\boldsymbol{q}_{\delta}$ on a batch of sample transitions $\mathcal{B}$,
\begin{equation}
    \mathcal{L}(\boldsymbol{q}, \boldsymbol{q}_{\delta}) = \frac{1}{|\mathcal{B}|} \sum_{b \in \mathcal{B} }(\boldsymbol{q}_{b} - \boldsymbol{q}_{\delta_b})^2.
\end{equation}
Note that because $\boldsymbol{q}_{\delta}$ is a copy of $\boldsymbol{q}$ where only the $i$-th element is altered via the update rule in \cref{eq:q_update_dqn}, the difference between all other entries in those two vectors is zero. As Q-values are defined in terms of $(s, a)$-pairs, this approach does not naturally apply to environments with continuous action spaces. In this case, the continuous action space has to be binned into a discrete representation.

\section{Quantum Q-learning}
\label{sec:qrl}

\begin{figure}[h]
\includegraphics[width=0.4\textwidth]{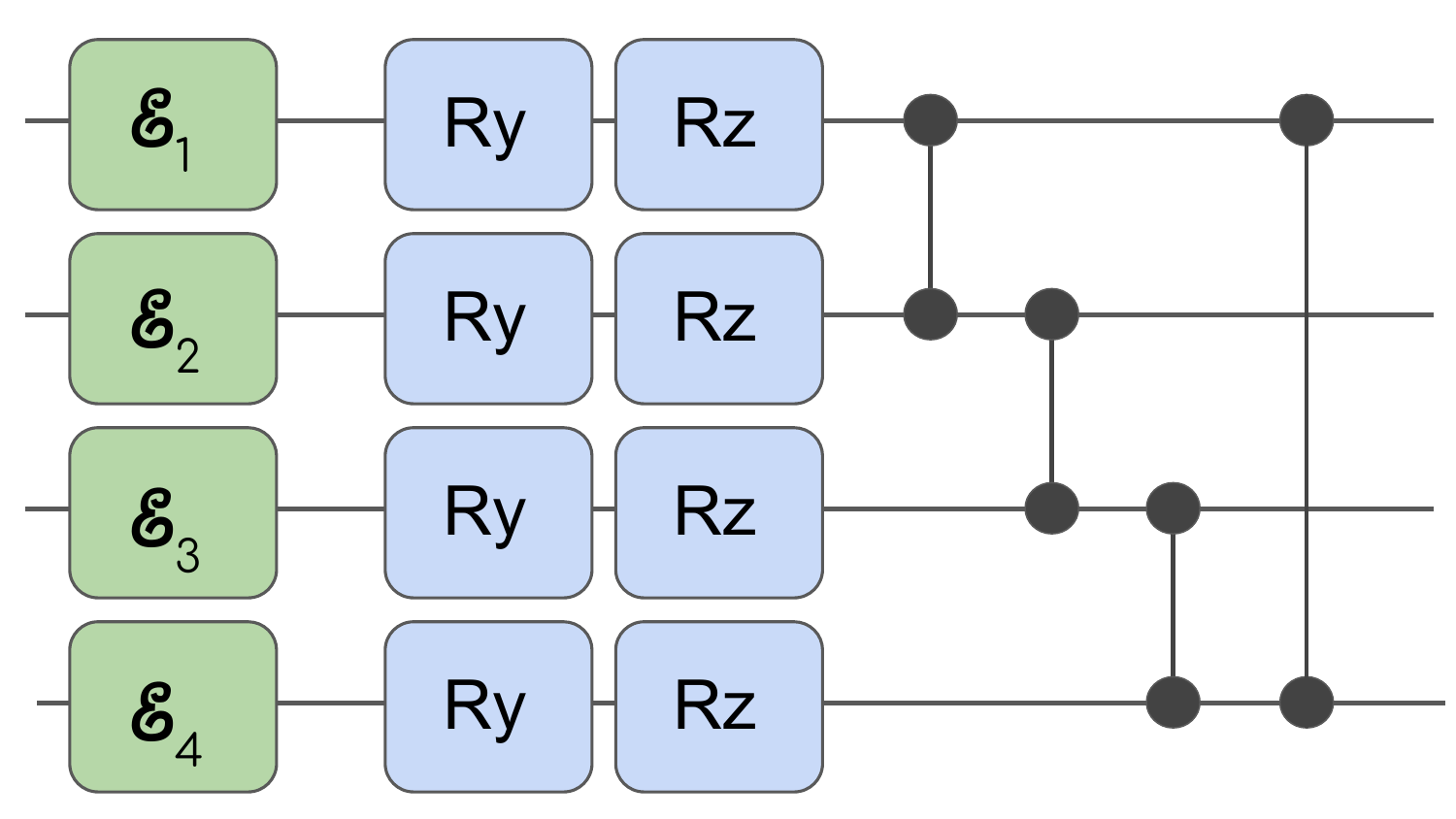}
\caption{PQC architecture used in this work. Each layer consists of a parametrized rotation along the $Y$ and $Z$ axes on each qubit, and a daisy chain of CZ gates. The green boxes correspond to data encoding gates that encode data as parameters of X rotations. When data re-uploading is used, the whole circuit pictured is repeated in each layer, without data re-uploading only the variational part without the initial X rotations is repeated.}
\label{fig:pqc}
\end{figure}

In this work, we adapt the DQN algorithm to use a PQC as its Q-function approximator instead of a NN. For this, we use a hardware-efficient ansatz \cite{kandala2017hardware} as shown in \cref{fig:pqc}. This ansatz is known to be highly expressive, and is susceptible to the barren plateau phenomenon for a large number of qubits and layers, although this is not an issue for the small state and action spaces we consider here. All other aspects of the Q-learning algorithm described in \cref{sec:q-learning} stay the same: we use a target network, an $\epsilon$-greedy policy to determine the agent's next action, and experience replay to draw samples for training the Q-network PQC. Our Q-network PQC is then $U_{\boldsymbol{\theta}}(s)$ parametrized by $\boldsymbol{\theta}$ and the target network PQC is $\hat{U}_{\boldsymbol{\theta}_{\delta}}(s)$, where $\boldsymbol{\theta}_{\delta}$ is a snapshot of the parameters $\boldsymbol{\theta}$ which is taken after fixed intervals of episodes $\delta$ and the circuit is otherwise identical to that of $U_{\boldsymbol{\theta}}(s)$. We now explain how environment states are encoded into our quantum model, and how measurements are performed to obtain Q-values.

\subsection{Encoding environment states}
\label{sec:encoding}

Depending on the state space of the environment, we distinguish between two different types of encoding in this work:

\textit{Discrete state space}: Discrete states are mapped to bitstrings and then input into the model, where on an all-zero state the bits corresponding to ones in the input state are flipped.

\textit{Continuous state space}: For continuous input states, we scale each component $x$ of an input state vector $\boldsymbol{x}$ to $x' = \arctan(x) \in [-\pi/2, \pi/2]$ and then perform a variational encoding, which consists of rotations in the $X$ direction by the angles $x'$.

As shown in \cite{schuld2020effect}, when data is encoded into a PQC by local rotation gates along the $X$-axis, the PQC can only model simple sine functions of its input. To further increase the expressivity of the circuit, the data encoding can be repeated in two ways: either in parallel by increasing the number of qubits and duplicating the data encoding on them, or in sequence in an alternating fashion with the variational layers of the circuit. The latter is also referred to as \textit{data re-uploading} in \cite{perez2020data}. Where needed, we will introduce data re-uploading to our model in \cref{sec:numerics}.

The formalism introduced in \cite{schuld2020effect} establishes a connection between PQCs and partial Fourier series by showing that the functions a given PQC can model can be represented as a Fourier series, where the accessible frequency spectrum depends on the eigenvalues of the data encoding gates, and the coefficients depend on the architecture of the variational part of the PQC and the observable that defines the readout operation. They show that in models as ours, where data is encoded in form of Pauli rotations, only Fourier series up to a certain degree can be learned, where the degree depends on the number of times the encoding gate is repeated. Additionally, the scale of the input data must match the scale of the frequencies of the modeled function for the model to fit the target function exactly. Making the scaling of input data itself trainable to increase a PQC's expressivity has been suggested in \cite{perez2020data,jerbi2021variational}, which we will also use by introducing a weight $\boldsymbol{w}_d$ on the input data. The input value $x'_i$ then becomes:

\begin{equation}
    x'_i = \arctan(x_i \cdot w_{d_i})~,
\end{equation}
where $w_{d_i}$ is the weight for input $x_i$. We will illustrate the advantage of these enhanced data-encoding strategies numerically in \cref{sec:numerics}.

\subsection{Computing Q-values}
\label{sec:computing_qs}

The Q-values of our quantum agent are computed as the expectation values of a PQC that is fed a state $s$ as

\begin{equation}\label{eq:q_val_readout}
    Q(s, a) = \bra{0^{\otimes n}}U_{\boldsymbol{\theta}}^{\dagger}(s)O_aU_{\boldsymbol{\theta}}(s)\ket{0^{\otimes n}},
\end{equation}
where $O_a$ is an observable and $n$ the number of qubits, and our model outputs a vector including Q-values for each possible $O_a$ as described in \cref{sec:q-learning}. The type of measurements we perform to estimate Q-values will be described in more detail in \cref{sec:numerics} for each environment. Before that, we want to highlight why the way Q-values are read out from the PQC is an important factor that determines the success at solving the environment at hand. A key difference between PQCs and NNs is that a PQC has a fixed range of output defined by its measurements, while a NN's range of output values can change arbitrarily during training depending on its weights and activation function. To understand why this is an important difference in a RL setting, we need to recall that Q-values are an estimate of the expected return
\begin{equation*}
    \begin{split}
        Q_{\pi}(s, a) = \mathbb{E}_{\pi}[G_t|s_t = s, a_t = a]\\
        = \mathbb{E}_{\pi}\left[\sum_{k=0}^{H-1} \gamma^k r_{t+k+1}|s_t = s, a_t = a\right].
    \end{split}
\end{equation*}
This quantity is directly linked to the performance of the agent in a given environment, so the model needs to have the ability to match the range of optimal Q-values in order to approximate the optimal Q-function. This means that the observables in a PQC-based Q-learning agent need to be chosen with care, and highly depend on the specific environment. To provide a simple example where an insufficient range prevents an agent from solving an environment, consider tabular learning in an environment that consists of a single state $s$ and two actions $a_1$ and $a_2$, where the agent should learn to always pick $a_1$. One episode has a maximum length of $H=10$ when the agent picks $a_1$ in each time step, and otherwise terminates when the agents picks action $a_2$. We consider a modification where the values in the Q-table are capped at 1, i.e., Q-values can not become larger than one, and both Q-values are initialized at zero. The environment is such that the reward for each action is 1 and the Q-value corresponding to the optimal action is $> 1$. For simplicity we set $\alpha = 1$ and $\gamma = 1$, which gives us an optimal value $Q_*(s, a_1) = 10$. We now perform an update on both Q-values according to the update rule in \cref{eq:q_learning_update},
\[Q(s_t, a_t) \leftarrow r_{t+1} + \mathrm{argmax}_a Q(s_{t+1}, a).\]
For action $a_2$, the transition from $s$ leads to episode termination, so the update rule yields $Q(s, a_2) = r_{t+1} = 1$. For action $a_1$, we get $Q(s, a_1) = 2$, however, due to the capped Q-table, we also get $Q(s, a_1) = 1$ for this state-action pair. We see that after a single update according to this update rule, both Q-values will be one and due to the capped range of the Q-table the Q-values are already saturated. No further update can change the Q-values, which means that the agent can do no better than random guessing hereafter. This simple example illustrates why it is essential in a tabular Q-learning setting that the range of values in the Q-table accommodates the magnitude of optimal Q-values. Updates in the function approximation case like in the gradient-based DQN algorithm are more complex due to the regression task that the agent solves to perform parameter updates, however, a similar saturation can still occur as the update rule for Q-values is the same (see \cref{eq:q_update_dqn}).

We have seen that it is crucial for a PQC-based Q-learning agent to have an output range that matches that of the optimal Q-values that it seeks to approximate. There are two ways to approach this issue: (i) multiply PQC outputs by a fixed factor to increase their range in a way that accommodates the theoretical maximum Q-value, (ii) make the output range itself a trainable model parameter. Multiplying the outputs of the PQC by a fixed factor increases the range of output values, but at the cost of potentially being close to the estimated maximum from the beginning, which makes this approach more sensitive to randomness in model parameter initialization. In particular, as Q-values are initialized randomly depending on the initial parameters of the PQC, the Q-values for actions of a specific state might have large differences. Considering that the reward which controls the magnitude of change given by the Q-value updates in \cref{eq:q_update_dqn} is comparatively small and actions are picked based on the argmax policy $\mathrm{argmax}_a Q(s, a)$, it may take a long time before subsequent updates of Q-values will lead to the agent picking the right actions. Even if we consider models that are initialized such that all Q-values are close to zero in the beginning, the actual changes in the rotation angles that the PQC needs to perform for Q-values of large ranges can become very small. Especially on NISQ devices, these changes might be impractically small to be reliably performed and measured on hardware. For these reasons, we focus on option (ii). We add a trainable weight $w_o \in \mathbb{R}$ to each readout operation, so that the output Q-value $Q(s, a)$ becomes
\begin{equation}
    Q(s, a) = \bra{0^{\otimes n}}U_{\boldsymbol{\theta}}(s)^{\dagger}O_a U_{\boldsymbol{\theta}}(s)\ket{0^{\otimes n}} \cdot w_{o_a},
\end{equation}
and each action has a separate weight $w_{o_a}$. We make the weights multiplicative in analogy to weights in a NN. This gives the model the possibility to flexibly increase the magnitude of Q-values to match the given environment. Notably, the number of actions in an environment is usually small compared to the number of parameters in the model, so adding one extra weight corresponding to each action does not designate a large overhead. In \cref{sec:Cart Pole}, we numerically show that the approach of using a trainable weight on the output value outperforms multiplying the model output by a fixed factor that is motivated by the range of optimal Q-values.

\section{Separation between quantum and classical Q-learning in restricted environments}
\label{sec:separation}

In this section, we make formal statements about a separation between quantum and classical models for Q-learning in a restricted family of environments. These statements are based on recent results in supervised \cite{liu2021rigorous} and policy gradient based reinforcement learning \cite{jerbi2021variational}. The latter work constructs families of environments that are proven to be hard for any classical learner, but can be solved in polynomial time by a quantum learner in a policy learning setting. Learning policies is closely related to learning Q-values, however, Q-values contain more information about the environment per definition as they cover the whole state-action space. This means that it is not straightforward to generalize the results from \cite{jerbi2021variational} to a Q-learning setting. In this section, we will show under which conditions optimal Q-values can be inferred from optimal policies, so that the separation results in \cite{jerbi2021variational} also apply to the Q-learning case. The environments constructed in \cite{jerbi2021variational} are based on the supervised learning task introduced in \cite{liu2021rigorous}, which are proven to be classically hard assuming the widely-believed hardness of the discrete logarithm problem, but can be solved by a quantum learner in polynomial time. To understand how a separation in supervised learning can be generalized to a RL setting, it is important to state that any classification task can be turned into an environment for RL. To do this, rewards in the environment are assigned according to the prediction the agent makes. First examples of this were introduced in \cite{dunjko2017exponential} for cases where the environment allows quantum access to its states. A classification task like the one proposed in \cite{liu2021rigorous} can be turned into a RL task by simply assigning a reward of 1 (-1) for a correct (incorrect) classification, and defining an episode as being presented with a set of training samples. In this section, we will briefly revise the separation results for supervised learning given in \cite{liu2021rigorous} and those for policy gradient RL given in \cite{jerbi2021variational}, before we move on to characterize the types of environments that allow a generalization of the results in \cite{jerbi2021variational} to a Q-learning setting.

\subsection{A classification task based on the discrete logarithm problem}
\label{sec:dlp_classification}

The authors of \cite{liu2021rigorous} construct a classification task that is intractable for any classical learner, but can be solved by a quantum learner in polynomial time. The classification task is based on the discrete logarithm problem (DLP), and the separation relies on the the quantum learner's ability to perform the algorithm provided by Shor in \cite{shor1999polynomial} to solve the DLP efficiently.

\begin{definition}[Discrete logarithm problem]
Let $\mathbb{Z}_p^* = \{1, 2, \dots, p-1\}$ be the cyclic multiplicative group of integers modulo $p$ for a large prime $p$, and $g$ a generator of this group. The DLP is defined as computing $\log_g x$ for an input $x \in \mathbb{Z}_p^*$.
\end{definition}
It is widely believed that no classical algorithm can solve the DLP efficiently, however, it is proven that the algorithm provided by Shor can solve DLP in $\mathrm{poly}(n)$ time for $n = \lceil \log_2 p \rceil$ \cite{shor1999polynomial}.
Based on this, \cite{liu2021rigorous} construct a classification task with a concept class $\mathcal{C} = \{f_s\}_{s \in \mathbb{Z}_p^*}$ and data points defined over the data space $\mathcal{X} = \mathbb{Z}_p^* \subseteq \{0,1\}^n$ as

\begin{equation}
    f_s(x) = \begin{cases} \mbox{+1,} & \mbox{if } \log_g x \in [s, s+\frac{p-3}{2}] \\ \mbox{$-$1,} & \mbox{otherwise,} \end{cases} 
\end{equation}
where each concept $f_s: \mathbb{Z}_p^* \to \{-1, 1\}$ maps one half of the elements in $\mathbb{Z}_p^*$ to 1 and the other half to $-1$, which yields a linearly separable set of data points in log-space. A quantum learner can make use of the algorithm from \cite{shor1999polynomial} to compute the discrete logarithm and solve the resulting trivial learning task. However, if a classical learner could solve the above learning task this would imply that there exists an efficient classical algorithm that solves the DLP. This is contrary to the widely believed conjecture that no efficient classical algorithm can solve the DLP, and \cite{liu2021rigorous} proves that no classical learner can do better than random guessing.

To connect these results to the RL setting, it is useful to be a bit more precise and define some terminology. The learning task is defined as finding a decision rule $f^*$, which assigns a label $y \in \{-1, 1\}$ to data point $x \in \mathcal{X}$\footnote{Note that we are adhering to the notation given in \cite{liu2021rigorous}, where the asterisk stands for the learned decision rule and the function without an asterisk stands for the decision rule we seek to learn. This is the opposite of the notation used in Q-learning literature where $Q_*$ stands for the optimal Q-values, which we have followed in previous sections. The authors of \cite{jerbi2021variational} have also adopted the latter notation in their paper to describe the DLP classification task. We will stick to denoting the learned decision rule with an asterisk in this section.}. $f^*$ is learned on a set of labeled examples $S = \{x_i, y_i\}_{i=1, \dots, m}$ generated by the unknown decision rule, or \textit{concept}, $f$. An efficient learner needs to compute $f^*$ in time polynomial in $n$ that agrees with the labeling given by $f$ with high probability, or in other words reaches a high \textit{test accuracy} on unseen samples,

\begin{equation}\label{eq:accuracy}
    \mathrm{acc}_f(f^*) = \underset{x \in \mathcal{X}}{\mathrm{Pr}}[f(x) = f^*(x)].
\end{equation}
The authors of \cite{liu2021rigorous} prove that no efficient classical learner can achieve 
\[\mathrm{acc}_f(f^*) = \frac{1}{2}+\frac{1}{\mathrm{poly}(n)}~\] unless an efficient classical algorithm that solves the DLP exists, while there exists a quantum learner that achieves close to perfect accuracy with high probability in polynomial time.

\subsection{Learning optimal policies in environments based on the DLP classification task}
\label{sec:dlp_policies}

After stating the classification task based on the DLP in the previous section, we now briefly review how the authors of \cite{jerbi2021variational} construct families of environments based on the DLP classification task to transfer the separation results to RL. They show that (i) solving these environments is classically hard for any learner unless there exists an efficient classical algorithm that solves the DLP, (ii) there exists a quantum learner that can solve these environments in polynomial time. To understand how the DLP classification task can be used to construct a classically hard to solve RL environment, it is important to note again that any classification task can be trivially turned into a RL task by letting each data point $x \in \mathcal{X}$ denote a state in the environment, and giving rewards to the agent depending on whether it correctly assigns a state to its predefined label $y$. The rewards for the DLP classification task are 1 ($-1$) for a correct (false) classification. While \cite{liu2021rigorous} are interested in achieving a high test accuracy, in a RL setting we want to find an agent with close-to-optimal performance in the given environment. The authors of  \cite{jerbi2021variational} measure this performance in terms of a value function $V_{\pi}(s)$ for policy $\pi$ and state $s$,
\begin{equation}
    V_{\pi}(s) = \mathbb{E}_{\pi}\left[\sum_{t=0}^{H-1} \gamma^t r_{t}|s_t = s\right]
\end{equation}
which is the expected reward for following policy $\pi$ for an episode of length $H$ in state $s$. Based on the DLP classification task from \cite{liu2021rigorous}, the authors of \cite{jerbi2021variational} define three different environments that are classically hard to learn, where the value function of each of these environments is closely related to the accuracy in \cref{eq:accuracy} of the policy on the classification task. This allows them to get bounds on the value function as a function of bounds on the accuracy. Roughly speaking, by Theorem 1 of \cite{liu2021rigorous} no classical learner can achieve performance better than that of random guessing in $\mathrm{poly}(n)$ time on those environments, unless an efficient classical algorithm to solve the DLP exists. We will briefly explain the set-up of the quantum learner in \cite{jerbi2021variational}, before going into more detail on one of the families of environments they construct to show a separation between classical and quantum learners for policy learning.

A RL agent can be trivially constructed from the classifier in \cite{liu2021rigorous}, which is based on a classical support vector machine (SVM) that takes the samples that have been ``decrypted" by a quantum feature map as an input. (This type of classifier is also referred to as an \textit{implicit} SVM). However, to get a learner that more closely matches the parametrized training of a quantum learner done in \cite{jerbi2021variational}, they use a model where the feature embedding and classification task are both solved by a PQC. This method is referred to as an \textit{explicit} SVM. The explicit SVM comprises a feature-encoding unitary $U(x)$ applied on the all zero state, which they refer to as $\ket{\phi(x)} = U(x)\ket{0^{\otimes n}}$, a variational part $V(\boldsymbol{\theta})$ with parameters $\boldsymbol{\theta}$, and an observable $O$. The feature-encoding unitary for the DLP task is the same as used in \cite{liu2021rigorous} so that feature states take the following form for $k=n - t \log n$ for a constant $t$ related to noisy classification (we refer the reader to \cite{jerbi2021variational} for a detailed description of classification under noise),
\begin{equation}\label{eq:shor_feature_map}
    \ket{\phi(x)} = \frac{1}{\sqrt{2^k}} \sum_{i=0}^{2^k-1}\ket{x \cdot g^i}.
\end{equation}
These states can be efficiently prepared on a fault-tolerant quantum computer by a circuit that uses the algorithm proposed by Shor in \cite{shor1999polynomial} as a subroutine. 
It was proven in \cite{liu2021rigorous} that for all concepts $f_s$ the data points with labels 1 and $-1$, respectively, can be separated by a hyperplane with a large margin, and that this hyperplane always exists. The learning task of the PQC $V(\boldsymbol{\theta})$ is then to find this hyperplane. The hyperplanes are normal to states of the form 
\begin{equation}
    \ket{\phi_{s'}} = \frac{1}{\sqrt{(p-1)/2}} \sum_{i=0}^{(p-3)/2} \ket{g^{s'+i}},
\end{equation}
for $s' \in \mathbb{Z}_p^*$. A classifier $h_{s'}(x)$ for these data points can then be defined as
\begin{equation}
    h_{s'}(x) = \begin{cases} \mbox{1,} & \mbox{if } |\braket{\phi(x)|\phi_{s'}}|^2 / \Delta \geq 1/2 \\ \mbox{$-$1,} & \mbox{otherwise,} \end{cases} 
\end{equation}
where $\Delta = \frac{2^k+1}{p-1}$ is the largest value the inner product $|\braket{\phi(x)|\phi_{s'}}|^2$ takes and is used to renormalize it to $[0, 1]$. The variational circuit is defined as $V(\boldsymbol{\theta}) = \hat{V}(s')$ which is similar in implementation to $U(x_i)$ with $x_i = g^{s'}$ and $k \approx n/2$, and a measurement operator $O = \ket{0^{\otimes n}}\bra{0^{\otimes n}}$.

The simplest way of turning the DLP classification task into an environment is to define one episode as the agent being in a randomly chosen state corresponding to a training sample, performing an action which assigns the predicted label, and giving a reward of 1 (-1) for a correct (incorrect) classification. This family of environments is referred to as SL-DLP in \cite{jerbi2021variational}. While the family of SL-DLP environments is a straightforward way to generalize the results from \cite{liu2021rigorous} to policy learning, it lacks the characteristics typically associated with RL, namely a temporal structure in the state transitions, such that these depend on the actions taken by the agent. To construct a family of environments based on the DLP which includes this kind of structure, \cite{jerbi2021variational} introduce the family of Cliffwalk-DLP environments, inspired by the textbook Cliffwalk environment from \cite{sutton2018reinforcement}. Here, the goal is still to assign correct labels to given states, but now these states follow a randomly assigned but fixed order. The agent has to ``walk along the edge of a cliff", where this edge is represented by the sequence of ordered states the environment takes. A correct classification leads to the next state in the sequence, while an incorrect classification leads to ``falling off the cliff" and immediate episode termination. The authors of \cite{jerbi2021variational} show that the quantum learnability results of the SL-DLP environment also hold for the family of Cliffwalk-DLP environments. In the following section we will generalize these results to Q-values by giving a definition of the types of environments where knowledge of an optimal policy lets us infer optimal Q-values.

\subsection{Estimating optimal Q-values from optimal policies}

In \cref{sec:dlp_policies}, we revised how \cite{jerbi2021variational} construct an efficient quantum agent that can achieve close-to-optimal policies in families of environments based on the DLP. Now, we turn to generalizing their results to the Q-learning setting. The classical hardness of the environment still holds irrespective of the learner that is used. The remaining question is now whether there exists an efficient algorithm to obtain optimal Q-values, given we have access to an optimal policy. Concretely, our goal is to compute optimal Q-values $Q_*(s, a)$ for state-action pairs from an environment, where $s$ is given by the environment and $a$ is determined by the optimal policy. 

One could imagine that $Q_*(s, a)$ can be easily estimated using Monte Carlo sampling since the definition involves only the use of the optimal policy after the move $(s, a)$ (cf. \cref{eq:q_exp_return}). However, in general it is not possible for an agent to get to arbitrary states $s$ in poly time. We circumvent this problem by considering special cases of environments that are classically hard, where there are only two actions $\{a, a’\}$, and where the analytic values of $Q_*(s,a)$ and $Q_*(s,a')$ are known. The only unknown is which action $a$ or $a'$ is the optimal one. In this case it is clear that access to the optimal policy resolves the question.

As an example of such an environment, consider the SL-DLP family of environments from \cite{jerbi2021variational}. In each episode, the agent needs to classify one random sample from a set of samples corresponding to the DLP classification task from \cref{sec:dlp_classification}, where a correct (incorrect) classification yields a reward of 1 (-1). If we set $\gamma=0$, the two possible Q-values for a given state and the two possible actions are simply the rewards corresponding to the result of the classification. To get the Q-value $Q_*(s, a)$, we query the policy $\pi_*(a|s)$ for the optimal action and assign the reward for a correct classification to the corresponding Q-value. (Note that we can also directly infer $Q_*(s, a')$ for the wrong action $a'$ from this, as there are only two distinct Q-values.) This can also be trivially extended to episodes with a horizon greater than one and $\gamma > 0$. After querying the policy for the optimal action given the initial state of the episode, the expected return is computed directly assuming optimal actions until the end of the episode is reached. I.e., we simply compute
\[Q_*(s_t, a_t) = \mathbb{E}_{\pi_*}\left[\sum_{k=0}^{H-1} \gamma^k r_{t+k+1}|s_t = s, a_t = a\right]\]
for $a_t$ given by the optimal policy, where all rewards are one from time step $t$ onward. (For more details on settings with longer horizon and a discount factor larger than zero, and an analytic expression of the Q-values in these cases, see \cite{jerbi2021variational}).

In more general cases, the issue of approximation reduces to the problem of reaching the desired state $s$ efficiently.
When this is possible (i.e., it is possible to construct environments which allow this without becoming easy to learn), then so is estimating Q-values given an optimal policy. Note that for all of the above, the same caveat as in \cite{jerbi2021variational} applies, namely that this method of obtaining optimal Q-values does not resemble Q-learning in the sense that we use a tabular or DQN-type approach as shown in \cref{sec:q-learning}, and it is still an open question whether a rigorous quantum advantage can be shown in these settings for either policy-based RL or Q-learning.

\section{Numerical results}
\label{sec:numerics}

\begin{figure}
\includegraphics[width=0.5\textwidth]{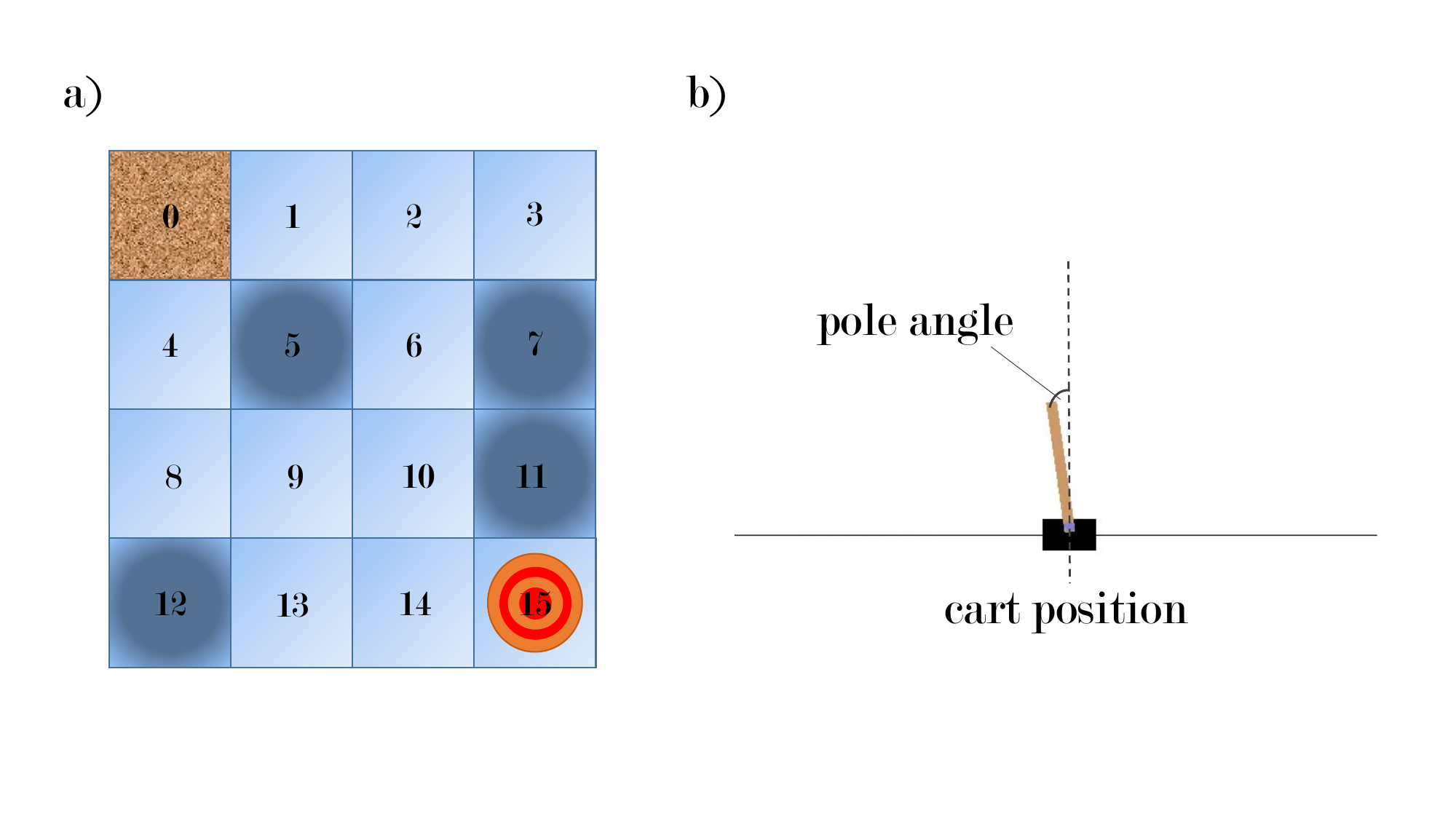}
\caption{Gym environments solved by the quantum model. a) Frozen Lake environment, where an agent needs to learn to navigate from the top left of a grid to retrieve the Frisbee at the bottom right without falling into any of the holes (dark squares), b) Cart Pole environment, which consists of learning to balance a pole on a cart which moves left and right on a frictionless track.}
\label{fig:envs}
\end{figure}

In this section, we present results for our PQC model on two benchmark RL tasks from the OpenAI Gym \cite{brockman2016openai}, Frozen Lake v0 \cite{fl_wiki} and Cart Pole v0 \cite{cp_wiki} (see \cref{fig:envs}). We ran an extensive hyperparameter search for both environments, and present our results for the best sets of hyperparameters. A detailed description of the hyperparameters we tuned and their best values can be found in \cref{sec:appendix_hps}. Our experiments were run with TensorFlow Quantum \cite{broughton2020tensorflow} and Cirq \cite{cirq}, the full code can be found on Github \cite{quantum_agents_code}.

\subsection{Frozen Lake}
\label{sec:frozen_lake}

\begin{figure*}
  \subfloat[average scores]{\includegraphics[scale=0.5]{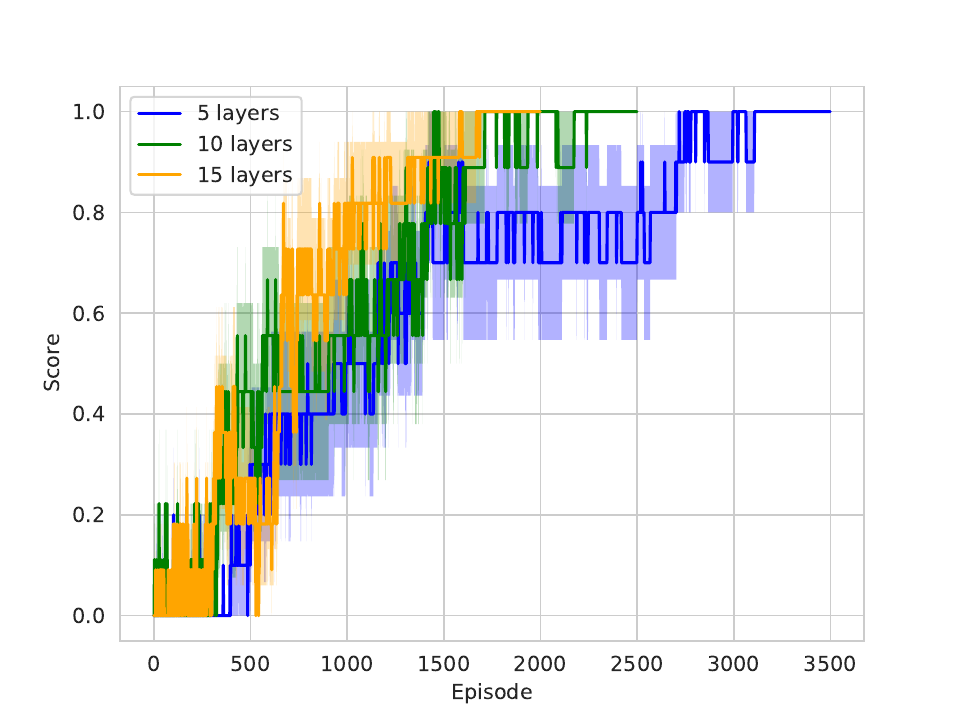}}\quad
  \subfloat[mean absolute error with optimal Q-values]{\includegraphics[scale=0.5]{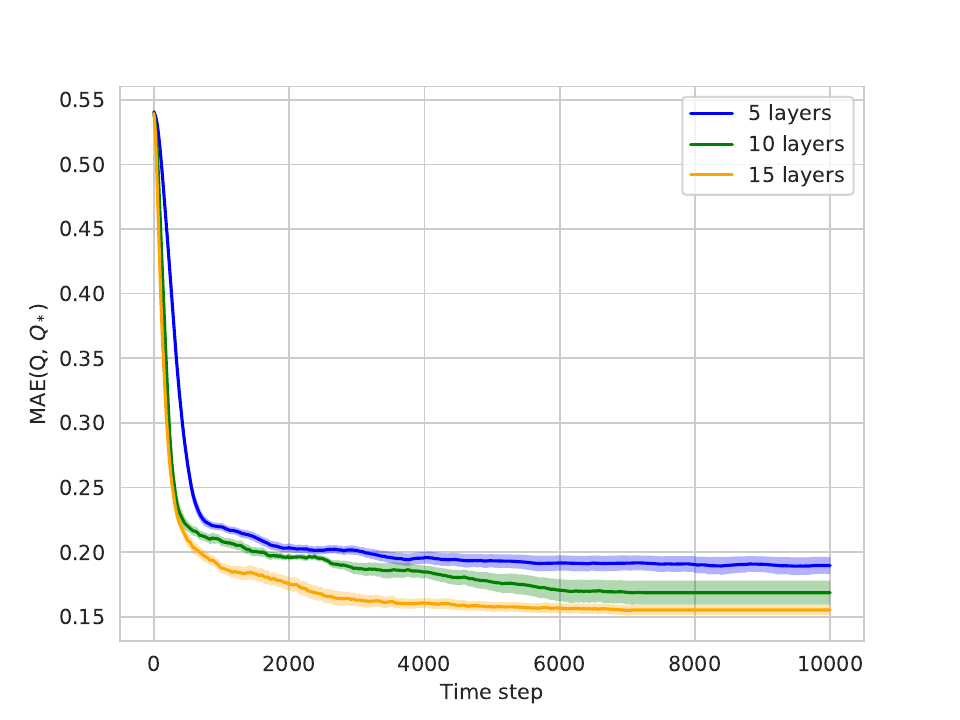}}
  \caption{Agents with varying depth playing the Frozen Lake environment, and their closeness to the optimal Q-values. The environment is solved when the agent reaches the goal (receives a score of 1) for 100 contiguous episodes. a) Average score over 10 agents for circuits of depth 5, 10, and 15, respectively. All agents manage to solve the environment, higher circuit depth leads to lower time to convergence. Shaded area shows standard deviation from the mean. b) Mean absolute error between agents' Q-values and the optimal Q-values $Q_*$ for all $(s, a)$ pairs over time steps in episodes, where one time step corresponds to one transition in the environment. Shaded region shows standard error of the mean.}
\label{fig:fl_results}
\end{figure*}

The Frozen Lake (FL) environment serves as an example for environments with a simple, discrete state space and with a reward structure that allows us to use an agent which performs measurements in the Z-basis to compute Q-values without the need for trainable weights to scale the output range. It consists of a 4x4 grid representing a frozen surface, where the agent can choose to move one step up, down, left or right. The goal is to cross the lake from the top left corner to the bottom right corner where the goal is located. However, some of the grid positions correspond to holes in the ice, and when the agent steps on them the episode terminates and it has to start again from the initial state. In each episode, the agent is allowed to take a maximum number of steps $m_{max}$. The episode terminates if one of the following conditions is met: the agent performs $m_{max}=200$ steps, reaches the goal, or falls into a hole. For each episode in which the goal is reached the agent receives a reward of 1, and a reward of 0 otherwise. The environment is considered solved when the agent reaches the goal for 100 contiguous episodes. (See \cite{fl_wiki} for full environment specification.)

As the FL environment is discrete and the dimensions of the state and action spaces are small, there is no true notion of generalization in this environment, as all distinct state-action pairs are likely observed during training. On the other hand, generalization to unseen state-action pairs is one of the key reasons why function approximation was introduced to Q-learning. For this reason, environments like Frozen Lake are not a natural fit for these types of algorithms and we refrain from comparing to a classical function approximator. Note that we also refrain from comparing to the tabular approach, as this is (i) guaranteed to converge and (ii) not interesting beyond environments with very limited state and action spaces. However, this environment is interesting from another perspective: there are only 64 Q-values which we can compute exactly, and therefore we can directly compare the Q-values learned by our model to the optimal Q-values $Q_*$, which is not possible for the continuous-state Cart Pole environment that we study in \cref{sec:Cart Pole}. We show the difference between our agents' Q-values and the optimal Q-values during the course of training in \cref{fig:fl_results} b). Additionally, the FL environment serves as a nice example for environments where a PQC with simple measurements in the $Z$-basis can be used to solve a RL task, without requiring additional post-processing, as we describe below.

The FL environment has 16 states (one for each square on the grid) of which four are holes (marked as darker squares in \cref{fig:envs} a), and 4 actions (top, down, left, right). We encode each position on the grid as one of the computational basis states of a 4-qubit system, without use of trainable input data weights or data re-uploading. The optimal Q-values for each state-action pair can be computed as $Q_*(s, a) = \gamma^{\beta}$ (cf. \cref{eq:q_exp_return}), where $\beta$ is the number of steps following the shortest path to the goal from the state $s'$ that the agent is in after the transition $(s, a)$. We will now motivate our choice of observables for the FL agent by studying the range the optimal Q-values can take. Note that these optimal Q-values are defined for the tabular case only, and serve as a reference for the Q-values we want our Q-function approximator to model. We know that only one transition, that from state 14 to the goal state 15, is rewarded. This corresponds to a Q-value $Q_*(14, R) = \gamma$. As the only other state adjacent to the goal (state 11) is a hole, no other transition in this environment is rewarded. Through the recursive Q-value update rule (see \cref{eq:q_learning_update}), all other Q-values depend on $Q_*(14, R)$, and are smaller due to the discount factor and the zero reward of all other transitions. In case of a function approximator, the Q-values may not be the same as the optimal values, but the relationship between $Q(14, R)$ and all other Q-values still applies as the update rule in \cref{eq:q_learning_update} changes values according to the observed reward and discounted expected reward. That is, if the function approximator outputs values that match the range of optimal Q-values and is not fundamentally limited in the updates that can be performed to it, the relationship above can be replicated. This means that we have an upper bound on the range of Q-values that we want to model which only depends on $\gamma \leq 1$ and stays constant over all episodes. Therefore we do not expect that Q-values need to become larger than $\gamma$ for our agent to solve the environment, and only become larger in practice if the initialization of our model happens to yield higher values for some state-action pairs. Motivated by this, we represent the Q-values for the four actions as the expectation values of a measurement with the operator $Z_i$ for each of the four qubits $i \in \{1, \dots, 4\}$, which we scale to lie between $[0, 1]$ instead of $[-1, 1]$. Note that even when parameter initialization yields Q-values higher than the largest optimal Q-value, they will still be close to this value as both optimal Q-values and those of our model are upper-bounded by 1.  
\Cref{fig:fl_results} a) shows the average scores of ten agents, each configuration trained with a circuit depth of 5, 10, and 15 layers, respectively. All agents manage to solve the environment, and the time to convergence decreases as the number of layers increases. \Cref{fig:fl_results} b) shows the averaged mean absolute error (MAE) between the optimal Q-values and the Q-values produced by the agents at each time step during training. The agents trained on circuits of depth 15 reach the lowest values and converge earlier to an average MAE that is roughly 0.05 lower than that of the agents trained on a circuit of depth 5. This illustrates that as we increase the complexity of the function approximator, the optimal Q-values can be more accurately modelled. However, the improvement between 10 and 15 layers is relatively small compared to that between 5 and 10 layers, similar to a saturation in performance w.r.t.\ number of parameters found in classical deep RL \cite{ota2021training}. We will study this type of scaling behaviour more in-depth and compare it to that of NNs in \cref{sec:Cart Pole}. At the same time, we see that producing optimal Q-values is not necessary to solve an environment, as we argue in \cref{sec:q-learning}. In the following section, we study an environment where we are not able to compute the optimal Q-values analytically due to the continuous state space, but where we compare to a classical approach to assess the quality of our solution instead.

\subsection{Cart Pole}
\label{sec:Cart Pole}

\begin{figure*}
  \subfloat[average scores with varying data encoding strategies for best set of hyperparameters]{\includegraphics[scale=0.5]{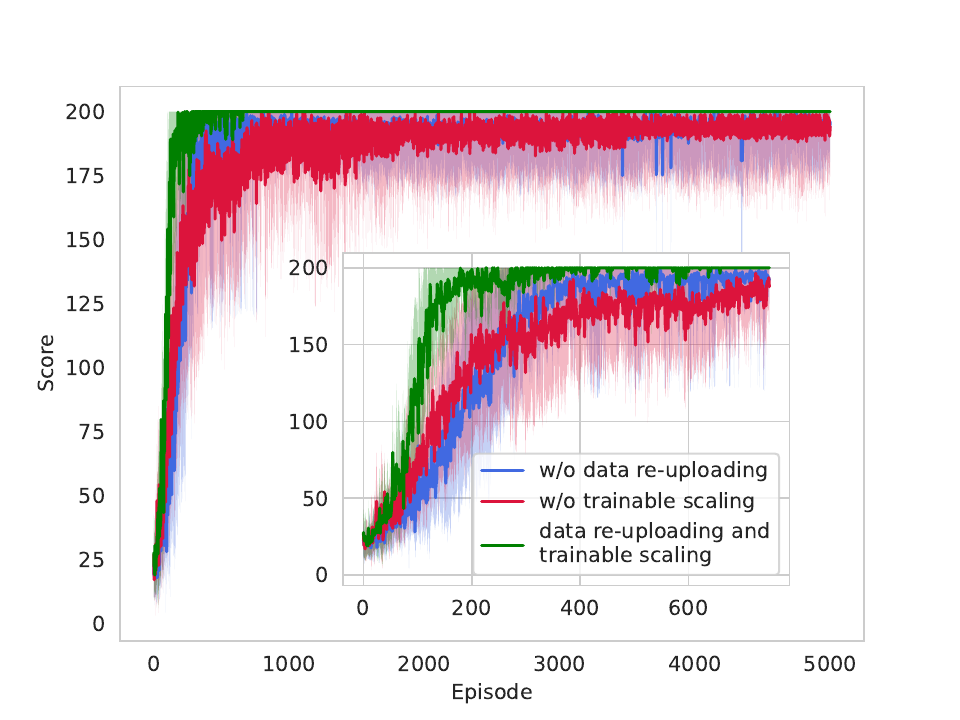}}\quad
  \subfloat[average scores with varying data encoding strategies for sub-optimal set of hyperparameters]{\includegraphics[scale=0.5]{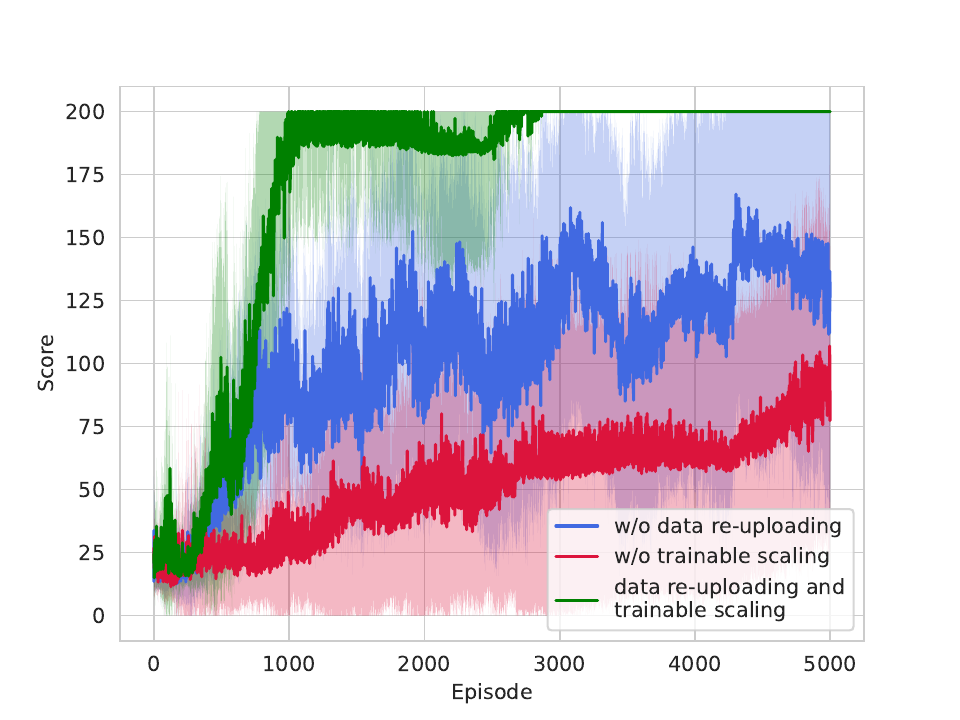}}
  \caption{Comparison of data-encoding strategies for the optimal and one sub-optimal set of hyperparameters for agents training in the Cart Pole environment. The environment is solved when an agent has an average reward $\geq195$ for the past 100 episodes, after which training is stopped. Results are averaged over 10 agents each, where each agent consists of 5 layers of the circuit architecture depicted in \cref{fig:pqc}.}
\label{fig:cp_results_dataenc}
\end{figure*}

In the previous section, we have seen that for an environment with discrete state space and a reward function that results in an upper bound of Q-values of one, a simple PQC without enhanced data encoding our readout strategies suffices to solve the environment. Now we turn to an environment that is slightly more complex: the continuous state space necessitates a more evolved data encoding strategy, while the reward function results in Q-values that far exceed the range of a Z-basis measurement. In the Cart Pole v0 environment, an agent needs to learn to balance a pole upright on a cart that moves on a frictionless track. The action space consists of two actions: moving the cart left and right. Its state space is continuous and consists of the following variables: cart position, cart velocity, pole angle, and pole velocity at the tip. The cart position is bounded between $\pm 2.4$, where values outside of this range mean leaving the space that the environment is defined in and terminating the episode. The pole angle is bounded between $\pm 41.8$°. The other two variables can take infinite values, but are bounded in practice by how episode termination is defined. An episode terminates if the pole angle is outside of $\pm 12$°, the cart position is outside of $\pm 2.4$, or the agent reaches the maximum steps per episode $m_{max}=200$. For each step of the episode (including the terminal step) the agent receives a reward of one. At the beginning of each episode, the four variables of the environment state are randomly initialized in a stable state within the range [-0.05, 0.05]. The episode score is computed as the cumulative reward of all steps taken in the episode. The environment is solved when the average score of the last 100 episodes is $\geq 195$. (See \cite{cp_wiki,sutton2018reinforcement} for full environment specification.)

As in \cref{sec:frozen_lake}, we now motive our choice of observables depending on how rewards are received in this environment. For this, we recall that a Q-value gives us the expected return for a given state-action pair,
\[Q_{\pi}(s, a) = \sum_{k=0}^{\infty} \gamma^k r_{t+k+1}. \]
Cart Pole is an episodic environment with a maximum number of time steps $H=200$ in the version of the environment we study here, so the Q-value following optimal policy $\pi_*$ from a stable state $s$ is
\[Q_*(s, a) = \sum_{k=0}^{H-1} \gamma^k. \]
When following an arbitrary policy $\pi$ and starting in a random stable state of the environment, the Q-value is
\[Q_{\pi}(s, a) = \sum_{k=0}^{h-1} \gamma^k, \]
where $h \leq H$ is the length of the episode which is determined by the policy. The longer the agent balances the pole, the higher $h$, with $h=H$ the maximum number of steps allowed in an episode. When not considering random actions taken by the $\epsilon$-greedy policy, $h$ depends solely on the performance of the agent, which changes as the agent gets better at balancing the pole. Consequently, the Q-values we want to approximate are lower bounded by the minimum number of steps it takes to make the episode terminate when always picking the wrong action (i.e., the pole doesn't immediately fall by taking one false action alone), and upper bounded by the Q-values assuming the optimal policy, where $h=H$. We stress that this upper bound applies to the optimal policy in one episode only, and that in practice the upper bound of the magnitude of Q-values during training depends on the performance of the agent as well as the number of episodes played. Compared to the range of expectation values of computational basis measurements these values can become very high, e.g.\ for $\gamma = 0.99$ we get $\mathrm{max}~Q_*(s, a) \approx 86$. Even when considering that Q-values need not necessarily be close to the optimal values to solve an environment, the range given by computational basis measurements is clearly too small compared to the frequency with which rewards are given and the number of episodes needed until convergence.

To give the agent the possibility to flexibly adjust it's output range, we add trainable weights on the output values as described in \cref{sec:computing_qs}. The Q-values now take the form 

\begin{equation}
    Q(s, a) = \frac{\bra{0^{\otimes 4}}U_{\boldsymbol{\theta}}(s)^{\dagger}O_aU_{\boldsymbol{\theta}}(s)\ket{0^{\otimes 4}}+1}{2} \cdot w_{o_a},
\end{equation}

where $O_{a=L}=Z_1Z_2$ and $O_{a=R}=Z_3Z_4$ are Pauli-$ZZ$ operators on qubits $(1, 2)$ and $(3, 4)$ respectively, corresponding to actions left and right. To further improve performance, we also use data re-uploading and add trainable weights on the input values as described in \cref{sec:encoding}.

\subsubsection{Comparison of data encoding and readout strategies}

\begin{figure*}
  \subfloat[average scores with varying output ranges for best set of hyperparameters]{\includegraphics[scale=0.5]{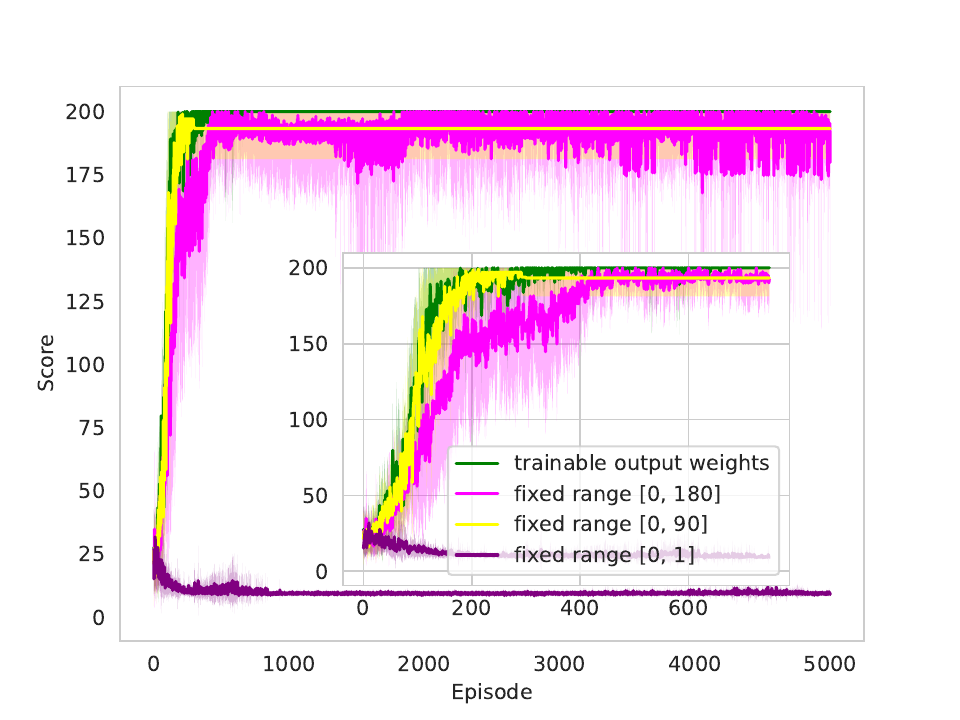}}\quad
  \subfloat[average scores with varying output ranges for sub-optimal set of hyperparameters]{\includegraphics[scale=0.5]{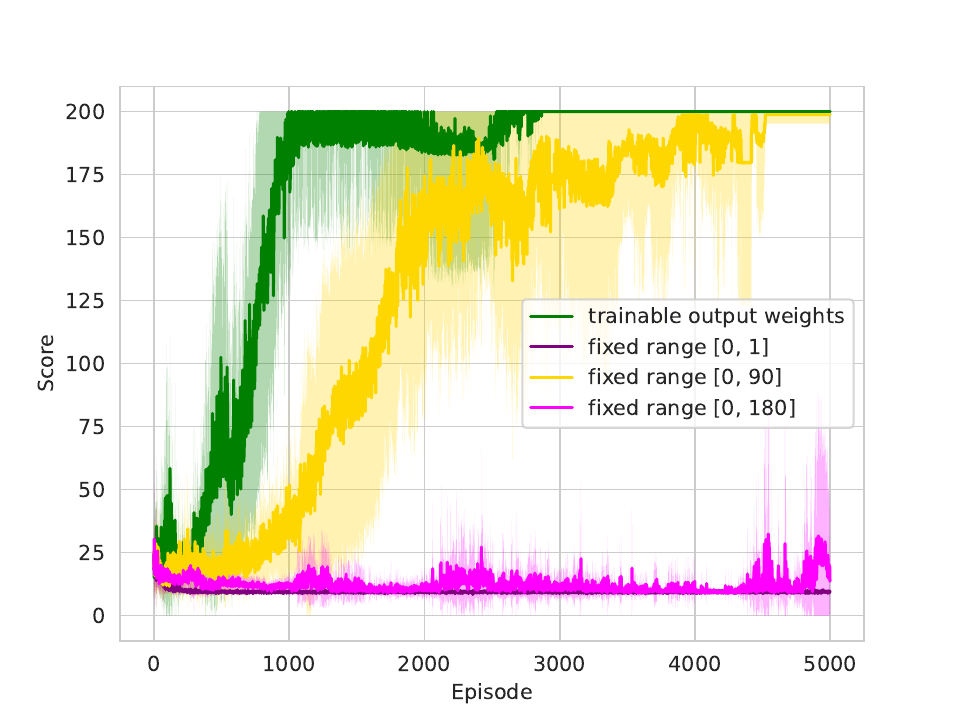}}
  \caption{Comparison of different readout strategies of the same agents as in \cref{fig:cp_results_dataenc} with the optimal and one sub-optimal set of hyperparameters.}
\label{fig:cp_results_readout}
\end{figure*}

To illustrate the effect of data re-uploading and trainable weights on the input and output values, we perform an ablation study and assess the impact of each of these enhancements on learning performance. To illustrate that our proposed architecture (i) performs better overall, and (ii) is less sensitive to changes in hyperparameters, we show results for the best set of hyperparameters that were found for a circuit of depth five, as well as a sub-optimal set of hyperparameters with which it is less easy for the agents to solve the Cart Pole environment. The hyperparameters we optimize over are: batch size, learning rates and update frequencies of the Q-value-generating model and the target model (cf. \cref{sec:q-learning}) (see \cref{sec:appendix_hps} for a detailed list of hyperparameter settings). Otherwise, we only vary the hyperparameters of the enhancements we want to study. The average performance of ten randomly initialized agents for each configuration is presented in \cref{fig:cp_results_dataenc} and \cref{fig:cp_results_readout}. Once an agent solves the environment, we stop training and in the figures show the last encountered score for each agent in the averages (i.e., to form averages over equal lengths of episodes, we assume that each agent continues scoring the same value as it did in its last interaction with the environment). 

\Cref{fig:cp_results_dataenc} a) and b) show the effects of varying data encoding strategies. While both data re-uploading and trainable weights on the input values alone do not produce agents that solve the environment in up to 5000 episodes for both the best and sub-optimal set of hyperparameters, combining both of these enhancements yields agents that solve Cart Pole in 3000 and 600 episodes at most on average, respectively. The fact that agents with trainable input weights and data re-uploading perform much better than those without, emphasizes the importance of matching the PQC's expressivity to the learning task at hand, as described in \cite{schuld2020effect}. In \cref{fig:cp_results_readout} a) and b), we compare agents with varying output ranges. Again, the green curves represent agents that are enhanced with a trainable weight corresponding to each Q-value that lets them flexibly adjust their output range during training, and these agents succeed with both sets of the remaining hyperparameters. The purple curves show agents with a fixed range of outputs of $[0, 1]$, all of which stay at an extremely low score during all 5000 episodes, as they fail to fit a good Q-function approximation regardless of hyperparameters. The yellow curves show agents with a fixed output range of maximally 90, which is motivated by the range of optimal Q-values. These agents also solve the environment on average, however, they are much more sensitive to parameter initialization and the remaining hyperparameters than agents with a trainable output range. The low final value of the yellow agents in \cref{fig:cp_results_readout} a) is due to their last interaction with the environment achieving a relatively low score on average.

As described above, the magnitude of Q-values crucially depends on the agent's ability to balance the pole in each episode, and as a general trend it will increase over the course of training for agents that perform well. How large the final Q-values of a solving agent are therefore also depends on the number of episodes it requires until convergence, so a range which is upper bounded by 90 presumes agents that converge relatively quickly. Considering the range of final Q-values of agents in the green curves, they can become as high as approximately 176 for agents that converge late. However, as we see for agents with a fixed output range of $[0, 180]$ (magenta curves), increasing the range to accommodate agents that converge later can lead to complete failure depending on the remaining hyperparameters.

\subsubsection{Comparison to the classical DQN algorithm}

\begin{figure*}
  \subfloat[PQCs, labels show: \#layers (\#parameters)]{\includegraphics[scale=0.5]{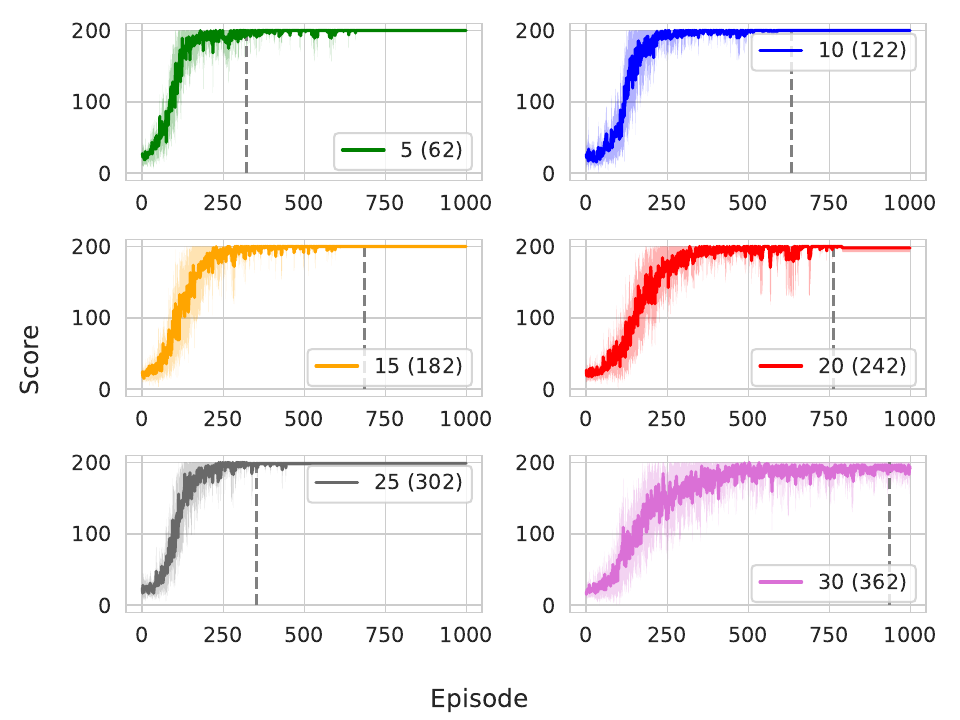}}\quad
  \subfloat[NNs, labels show: (\#units in hidden layer 1, 2), \#parameters]{\includegraphics[scale=0.5]{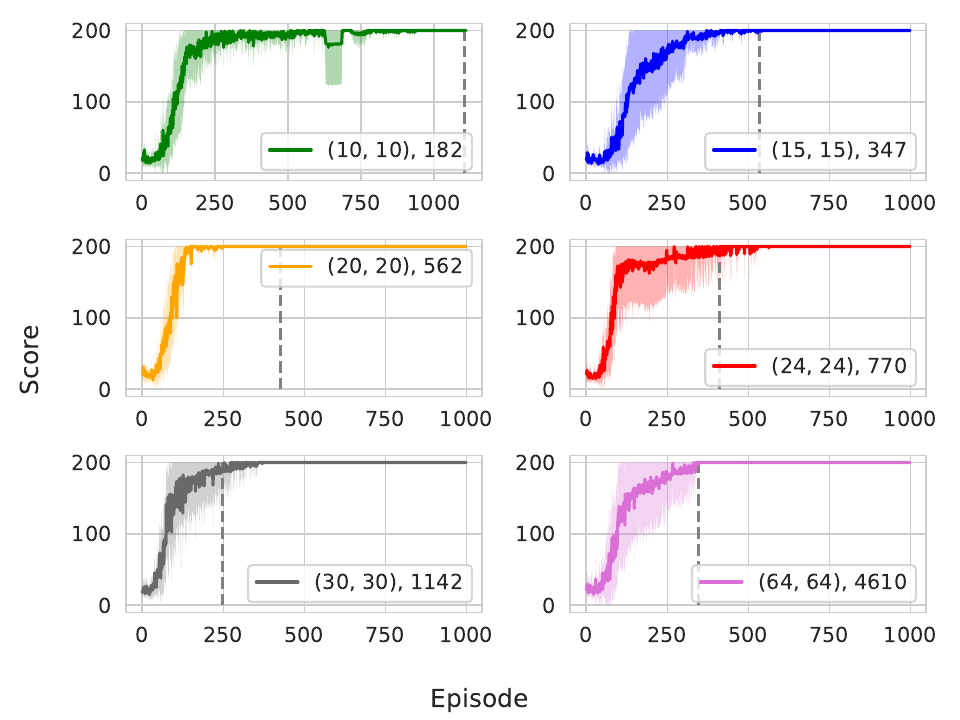}}
  \caption{Comparison of classical and quantum agents with varying numbers of parameters in the Cart Pole environment. Each sub-figure contains results averaged over ten agents, and the vertical dashed line marks the average number of episodes until solving the environment. We performed a hyperparameter optimization for each parameter configuration separately, and show the best setting for each. (See \cref{sec:appendix_hps} for all settings and a list of hyperparameters that were searched over.)}
\label{fig:cp_nn_comparison}
\end{figure*}
In addition to investigating the effects of varying data encoding and readout strategies, we compare the performance of our PQC model to that of the standard DQN algorithm that uses a NN as a function approximator. We do this for varying numbers of parameters for both the PQC and NN, and study how performance changes as the number of parameters increases. Note that because environments are strictly defined with a fixed number of input state variables, we cannot change the number of qubits arbitrarily for a certain environment. Studying varying system sizes in terms of qubits requires either artificially adjusting the data encoding to fit a certain number of qubits, or studying completely different environments all together. Therefore we focus on studying different model sizes in terms of number of parameters here. Additionally, the standard approach to increase model performance in supervised and unsupervised learning in the classical and quantum literature alike is often to add more parameters. However, it has been shown that this strategy does often not lead to success in classical deep RL due to the instability of training larger networks \cite{ota2021training}. Instead, it is much more important to find good settings of hyperparameters (including the random initialization of model parameters), and it is preferable to use models which are less sensitive to changes in these settings.

To study whether this effect is also present when the function approximator is a PQC, we compare agents with up to 30 layers of the hardware efficient ansatz depicted in \cref{fig:pqc}. All agents use the enhancements which have shown to yield good performance in \cref{fig:cp_results_dataenc} and \cref{fig:cp_results_readout}, namely data re-uploading and trainable input and output weights. The other hyperparameters that yield to the best performance for each depth are found through an extensive hyperparameter search and include the three different learning rates (Q-network, input and output weights), batch size, and update frequency of the Q-network and target network (see \cref{sec:appendix_hps} for detailed settings). \Cref{fig:cp_nn_comparison} a) shows the average performance over 10 quantum agents of each configuration.  We indeed observe that increasing the number of parameters is only efficient up to a certain point, after which additional layers lead to slower convergence. The best-performing configuration on average is a PQC with 25 layers and 302 parameters, which takes 500 episodes on average to solve the Cart Pole environment. 

\begin{figure}[h]
\includegraphics[width=0.5\textwidth]{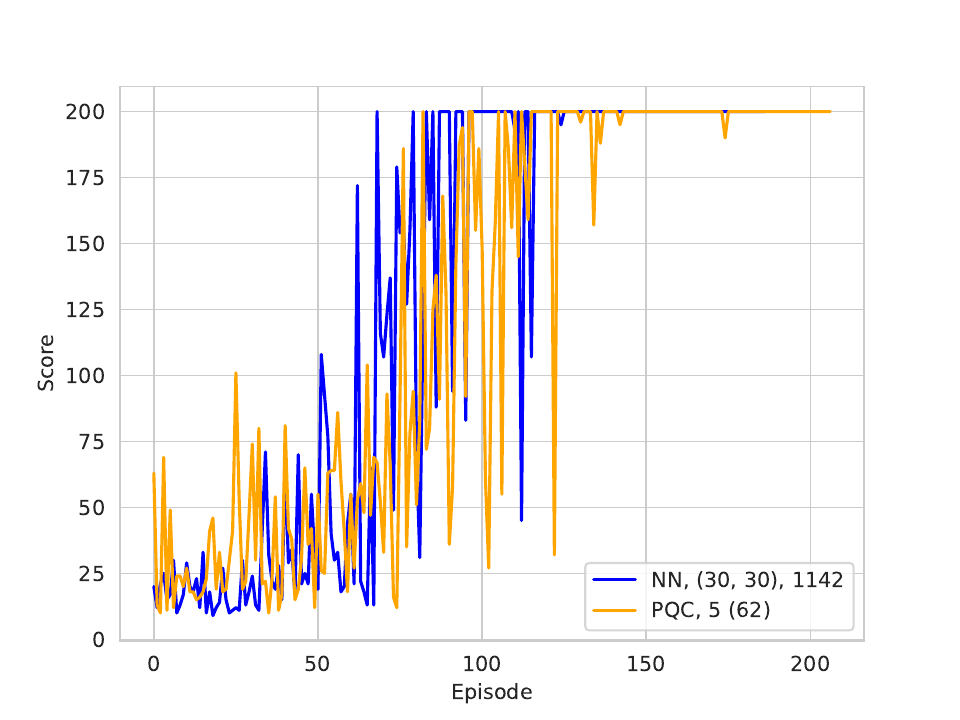}
\caption{Best PQC and NN from the configurations we study in \cref{fig:cp_nn_comparison}. The best PQC (orange, 5 layers, 62 parameters) takes 20 episodes longer to solve Cart Pole than the best NN (blue, two hidden layers with 30 units each, 1142 parameters).}
\label{fig:best_pqc_nn}
\end{figure}

To investigate the performance of the classical DQN algorithm which uses a NN as the function approximator, we compare NNs with two hidden layers with varying numbers of units. As simply increasing the depth of the NNs has not been beneficial in a RL setting, it has been proposed to use shallow networks with increased width instead \cite{ota2021training}. Therefore we keep the depth of our NNs fixed at two, and vary the width by changing the number of units in each hidden layer. This configuration is also inspired by well-performing agents on the official OpenAi Gym leaderboard \cite{gym_leaderboard}.\footnote{However, we note that it is hard to find reliable benchmarks on the Cart Pole environment in classical literature, as it was already too small to be considered in state-of-the-art deep learning when the DQN algorithm was introduced in \cite{mnih2015human}.} We make the same observation for the NNs in \cref{fig:cp_nn_comparison} b) as we did for the PQCs -- increasing the number of parameters does not necessarily improve performance. The best-performing NN is one with 20 units in each of its hidden layers, which yields a network with 562 parameters overall that solves the Cart Pole environment in 250 episodes on average. Comparing the configurations of PQC and NN that perform best on average, the best NN configuration takes roughly half as many episodes on average to solve Cart Pole than the best PQC, and does this with roughly twice as many parameters. Notably, the PQCs seem to suffer more from an instability during training as the number of parameters is increased than the NNs do. We also show a comparison of the best individual (not averaged) PQC and NN agents in \cref{fig:best_pqc_nn}. Here, the gap is relatively small: the best PQC (5 layers, 62 parameters) takes 206 episodes to solve Cart Pole, while the best NN (2 hidden layers with 30 units each, 1142 parameters) takes 186 episodes.

Finally, we note that unlike for the Frozen Lake environment, it is not straightforward to compute optimal Q-values for Cart Pole as its state space is continuous. A trained model that is known to implement the optimal policy (i.e., correct ordering of Q-values for all $(s,a)$-pairs) could be used as a baseline to compare other models to, but the magnitudes of Q-values can highly vary even among agents that solve the environment so this comparison will not provide much insight, which is why we refrain from including it here. Nonetheless, we provide a visualization of the Q-values learned by one of our best-performing quantum models in \cref{sec:appendix_vis}. We observe that these Q-values have a maximum value close to what we expected from an optimal agent (i.e., $86$).

\section{Conclusion}
\label{sec:conclusion}

In this work, we have proposed a quantum model for deep Q-learning which can encode states of environments with discrete and continuous state spaces. We have illustrated the importance of picking the observables of a quantum model such that it can represent the range of optimal Q-values that this algorithm should learn to approximate. One crucial difference between PQCs and classical methods based on NNs, namely the former's restricted range of output values defined by its measurement operators, was identified as a major impediment to successfully perform Q-learning in certain types of environments. Based on the range of optimal Q-values, we illustrate how an informed choice can be made for the quantum model's observables. We also introduce trainable weights on the observables of our model to achieve a flexible range of output values as given by a NN and empirically show the benefit of this strategy on the Cart Pole environment by performing ablation studies. Our results show that a trainable output range can lead to better performance as well as lower sensitivity to the choice of hyperparameters and random initialization of parameters of the model. We also perform ablation studies on a number of data encoding techniques which enhance the expressivity of PQCs, namely data re-uploading \cite{perez2020data} and trainable weights on the input \cite{perez2020data,jerbi2021variational}. We show the benefit of combining both approaches in the Cart Pole environment, where any of the two encoding strategies on its own does not suffice to reliably solve the environment. Our results illustrate the importance of architectural choices for QML models, especially for a RL algorithm as Q-learning that has very specific demands on the range of output values the model can produce. 

Additionally, we investigate whether recent results in classical deep Q-learning also hold for PQC-based Q-learning, namely that increasing the number of parameters in a model might lead to lower performance due to instability in training. To evaluate the performance of our model compared to the classical approach where the same DQN algorithm is used with a NN as the Q-function approximator, we study the performance of a number of classical and quantum models with increasing numbers of parameters. Our results confirm that PQC-based agents behave similarly to their NN counterparts as the number of parameters increases. Performance only increases up to a certain point and then declines afterward. We find that in both cases, the hyperparameter settings (and in case of the PQC data encoding and readout strategies) are the determining factors for a model's success much more than the number of parameters. This is in contrast to previous results for training PQCs on supervised and unsupervised learning tasks, where additional layers are likely to increase performance \cite{wiersema2020exploring,kiani2020learning,liu2018differentiable}. The effect that an increased number of parameters hampers performance in Q-learning also seems to be more prominent in PQCs than in NNs, which raises the question whether we need additional mechanisms to increase learning stability in this setting than the ones from classical literature.

In addition to our numerical studies, we also investigated whether a recent proof of quantum advantage for policy gradient RL agents \cite{jerbi2021variational} implies a separation of classical and quantum Q-learning agents as well. We show how optimal Q-values for state-action pairs can be efficiently computed given access to an optimal policy in the SL-DLP family of environments from \cite{jerbi2021variational}. We explain additional requirements on the structure of states in a given environment that need to be fulfilled to allow efficiently inferring optimal Q-values from optimal policies in more general environments. However, the separation results in \cite{jerbi2021variational} only guarantee that quantum learners can be constructed in general, and not that the optimal policy can be learned by policy gradient methods directly. It is an interesting open question if a separation between classical and quantum agents can also be proven for learning algorithms that use policy gradient or Q-value updates as shown in \cref{eq:q_learning_update}. This opens up the path to future investigations of possible quantum advantages of these types of quantum agents in relevant settings.

\section*{Acknowledgements}
AS thanks Casper Gyurik for many valuable discussions over the course of this project. This work was in part supported by the Dutch Research Council (NWO/OCW), as part of the Quantum Software Consortium program (project number 024.003.037). VD also acknowledges the support by the project NEASQC funded from the European Union’s Horizon 2020 research and innovation programme (grant agreement No 951821), and partial funding by an unrestricted gift from Google Quantum AI. SJ acknowledges support from the Austrian Science Fund (FWF) through the projects DK-ALM:W1259-N27 and SFB BeyondC F7102. SJ also acknowledges the Austrian Academy of Sciences as a recipient of the DOC Fellowship. AS acknowledges funding from the European Union's Horizon 2020 research and innovation programme under the Grant Agreement No. 828826.

\bibliographystyle{unsrturl}
\bibliography{lib}

\begin{thebibliography}{10}

\bibitem{bharti2021noisy}
Kishor Bharti, Alba Cervera-Lierta, Thi~Ha Kyaw, Tobias Haug, Sumner
  Alperin-Lea, Abhinav Anand, Matthias Degroote, Hermanni Heimonen, Jakob~S
  Kottmann, Tim Menke, et~al.
\newblock Noisy intermediate-scale quantum (nisq) algorithms.
\newblock {\em arXiv preprint arXiv:2101.08448}, 2021.

\bibitem{preskill2018quantum}
John Preskill.
\newblock Quantum computing in the nisq era and beyond.
\newblock {\em Quantum}, 2:79, 2018.
\newblock \href {https://doi.org/10.22331/q-2018-08-06-79}
  {\path{doi:10.22331/q-2018-08-06-79}}.

\bibitem{mitarai2018quantum}
Kosuke Mitarai, Makoto Negoro, Masahiro Kitagawa, and Keisuke Fujii.
\newblock Quantum circuit learning.
\newblock {\em Physical Review A}, 98(3):032309, 2018.
\newblock \href {https://doi.org/10.1103/PhysRevA.98.032309}
  {\path{doi:10.1103/PhysRevA.98.032309}}.

\bibitem{schuld2020circuit}
Maria Schuld, Alex Bocharov, Krysta~M Svore, and Nathan Wiebe.
\newblock Circuit-centric quantum classifiers.
\newblock {\em Physical Review A}, 101(3):032308, 2020.
\newblock \href {https://doi.org/10.1103/PhysRevA.101.032308}
  {\path{doi:10.1103/PhysRevA.101.032308}}.

\bibitem{schuld2019quantum}
Maria Schuld and Nathan Killoran.
\newblock Quantum machine learning in feature hilbert spaces.
\newblock {\em Physical review letters}, 122(4):040504, 2019.
\newblock \href {https://doi.org/10.1103/PhysRevLett.122.040504}
  {\path{doi:10.1103/PhysRevLett.122.040504}}.

\bibitem{havlivcek2019supervised}
Vojt{\v{e}}ch Havl{\'\i}{\v{c}}ek, Antonio~D C{\'o}rcoles, Kristan Temme,
  Aram~W Harrow, Abhinav Kandala, Jerry~M Chow, and Jay~M Gambetta.
\newblock Supervised learning with quantum-enhanced feature spaces.
\newblock {\em Nature}, 567(7747):209--212, 2019.
\newblock \href {https://doi.org/10.1038/s41586-019-0980-2}
  {\path{doi:10.1038/s41586-019-0980-2}}.

\bibitem{farhi2018classification}
Edward Farhi and Hartmut Neven.
\newblock Classification with quantum neural networks on near term processors.
\newblock {\em arXiv preprint arXiv:1802.06002}, 2018.

\bibitem{amin2018quantum}
Mohammad~H Amin, Evgeny Andriyash, Jason Rolfe, Bohdan Kulchytskyy, and Roger
  Melko.
\newblock Quantum boltzmann machine.
\newblock {\em Physical Review X}, 8(2):021050, 2018.
\newblock \href {https://doi.org/10.1103/PhysRevX.8.021050}
  {\path{doi:10.1103/PhysRevX.8.021050}}.

\bibitem{coyle2020born}
Brian Coyle, Daniel Mills, Vincent Danos, and Elham Kashefi.
\newblock The born supremacy: Quantum advantage and training of an ising born
  machine.
\newblock {\em npj Quantum Information}, 6(1):1--11, 2020.
\newblock \href {https://doi.org/10.1038/s41534-020-00288-9}
  {\path{doi:10.1038/s41534-020-00288-9}}.

\bibitem{zoufal2020variational}
Christa Zoufal, Aur{\'e}lien Lucchi, and Stefan Woerner.
\newblock Variational quantum boltzmann machines.
\newblock {\em Quantum Machine Intelligence}, 3(1):1--15, 2021.
\newblock \href {https://doi.org/10.1007/s42484-020-00033-7}
  {\path{doi:10.1007/s42484-020-00033-7}}.

\bibitem{lloyd2018quantum}
Seth Lloyd and Christian Weedbrook.
\newblock Quantum generative adversarial learning.
\newblock {\em Physical review letters}, 121(4):040502, 2018.
\newblock \href {https://doi.org/10.1103/PhysRevLett.121.040502}
  {\path{doi:10.1103/PhysRevLett.121.040502}}.

\bibitem{zoufal2019quantum}
Christa Zoufal, Aur{\'e}lien Lucchi, and Stefan Woerner.
\newblock Quantum generative adversarial networks for learning and loading
  random distributions.
\newblock {\em npj Quantum Information}, 5(1):1--9, 2019.
\newblock \href {https://doi.org/10.1038/s41534-019-0223-2}
  {\path{doi:10.1038/s41534-019-0223-2}}.

\bibitem{chakrabarti2019quantum}
Shouvanik Chakrabarti, Huang Yiming, Tongyang Li, Soheil Feizi, and Xiaodi Wu.
\newblock Quantum wasserstein generative adversarial networks.
\newblock In {\em Advances in Neural Information Processing Systems}, pages
  6781--6792, 2019.

\bibitem{hamann2020quantum}
A~Hamann, V~Dunjko, and S~W{\"o}lk.
\newblock Quantum-accessible reinforcement learning beyond strictly epochal
  environments.
\newblock {\em arXiv preprint arXiv:2008.01481}, 2020.
\newblock \href {https://doi.org/10.1007/s42484-021-00049-7}
  {\path{doi:10.1007/s42484-021-00049-7}}.

\bibitem{jerbi2019quantum}
Sofiene Jerbi, Lea~M Trenkwalder, Hendrik~Poulsen Nautrup, Hans~J Briegel, and
  Vedran Dunjko.
\newblock Quantum enhancements for deep reinforcement learning in large spaces.
\newblock {\em PRX Quantum}, 2(1):010328, 2021.
\newblock \href {https://doi.org/10.1103/PRXQuantum.2.010328}
  {\path{doi:10.1103/PRXQuantum.2.010328}}.

\bibitem{chen2020variational}
Samuel Yen-Chi Chen, Chao-Han~Huck Yang, Jun Qi, Pin-Yu Chen, Xiaoli Ma, and
  Hsi-Sheng Goan.
\newblock Variational quantum circuits for deep reinforcement learning.
\newblock {\em IEEE Access}, 8:141007--141024, 2020.
\newblock \href {https://doi.org/10.1109/ACCESS.2020.3010470}
  {\path{doi:10.1109/ACCESS.2020.3010470}}.

\bibitem{lockwood2020reinforcement}
Owen Lockwood and Mei Si.
\newblock Reinforcement learning with quantum variational circuit.
\newblock In {\em Proceedings of the AAAI Conference on Artificial Intelligence
  and Interactive Digital Entertainment}, pages 245--251, 2020.

\bibitem{wu2020quantum}
Shaojun Wu, Shan Jin, Dingding Wen, and Xiaoting Wang.
\newblock Quantum reinforcement learning in continuous action space.
\newblock {\em arXiv preprint arXiv:2012.10711}, 2020.

\bibitem{benedetti2019parameterized}
Marcello Benedetti, Erika Lloyd, Stefan Sack, and Mattia Fiorentini.
\newblock Parameterized quantum circuits as machine learning models.
\newblock {\em Quantum Science and Technology}, 4(4):043001, 2019.
\newblock \href {https://doi.org/10.1088/2058-9565/ab4eb5}
  {\path{doi:10.1088/2058-9565/ab4eb5}}.

\bibitem{jerbi2021variational}
Sofiene Jerbi, Casper Gyurik, Simon Marshall, Hans Briegel, and Vedran Dunjko.
\newblock Parametrized quantum policies for reinforcement learning.
\newblock {\em Advances in Neural Information Processing Systems}, 34, 2021.

\bibitem{mnih2015human}
Volodymyr Mnih, Koray Kavukcuoglu, David Silver, Andrei~A Rusu, Joel Veness,
  Marc~G Bellemare, Alex Graves, Martin Riedmiller, Andreas~K Fidjeland, Georg
  Ostrovski, et~al.
\newblock Human-level control through deep reinforcement learning.
\newblock {\em nature}, 518(7540):529--533, 2015.
\newblock \href {https://doi.org/10.1038/nature14236}
  {\path{doi:10.1038/nature14236}}.

\bibitem{silver2016mastering}
David Silver, Aja Huang, Chris~J Maddison, Arthur Guez, Laurent Sifre, George
  Van Den~Driessche, Julian Schrittwieser, Ioannis Antonoglou, Veda
  Panneershelvam, Marc Lanctot, et~al.
\newblock Mastering the game of go with deep neural networks and tree search.
\newblock {\em nature}, 529(7587):484--489, 2016.
\newblock \href {https://doi.org/10.1038/nature16961}
  {\path{doi:10.1038/nature16961}}.

\bibitem{berner2019dota}
Christopher Berner, Greg Brockman, Brooke Chan, Vicki Cheung, Przemyslaw
  Debiak, Christy Dennison, David Farhi, Quirin Fischer, Shariq Hashme, Chris
  Hesse, et~al.
\newblock Dota 2 with large scale deep reinforcement learning.
\newblock {\em arXiv preprint arXiv:1912.06680}, 2019.

\bibitem{vinyals2019grandmaster}
Oriol Vinyals, Igor Babuschkin, Wojciech~M Czarnecki, Micha{\"e}l Mathieu,
  Andrew Dudzik, Junyoung Chung, David~H Choi, Richard Powell, Timo Ewalds,
  Petko Georgiev, et~al.
\newblock Grandmaster level in starcraft ii using multi-agent reinforcement
  learning.
\newblock {\em Nature}, 575(7782):350--354, 2019.
\newblock \href {https://doi.org/10.1038/s41586-019-1724-z}
  {\path{doi:10.1038/s41586-019-1724-z}}.

\bibitem{konda2000actor}
Vijay~R Konda and John~N Tsitsiklis.
\newblock Actor-critic algorithms.
\newblock In {\em Advances in neural information processing systems}, pages
  1008--1014, 2000.

\bibitem{mnih2016asynchronous}
Volodymyr Mnih, Adria~Puigdomenech Badia, Mehdi Mirza, Alex Graves, Timothy
  Lillicrap, Tim Harley, David Silver, and Koray Kavukcuoglu.
\newblock Asynchronous methods for deep reinforcement learning.
\newblock In {\em International conference on machine learning}, pages
  1928--1937. PMLR, 2016.

\bibitem{watkins1989learning}
Christopher John Cornish~Hellaby Watkins.
\newblock Learning from delayed rewards.
\newblock 1989.

\bibitem{smith2018disciplined}
Leslie~N Smith.
\newblock A disciplined approach to neural network hyper-parameters: Part
  1--learning rate, batch size, momentum, and weight decay.
\newblock {\em arXiv preprint arXiv:1803.09820}, 2018.

\bibitem{ye2019comparison}
Ziyu Ye, Andrew Gilman, Qihang Peng, Kelly Levick, Pamela Cosman, and Larry
  Milstein.
\newblock Comparison of neural network architectures for spectrum sensing.
\newblock In {\em 2019 IEEE Globecom Workshops (GC Wkshps)}, pages 1--6. IEEE,
  2019.
\newblock \href {https://doi.org/10.1109/GCWkshps45667.2019.9024482}
  {\path{doi:10.1109/GCWkshps45667.2019.9024482}}.

\bibitem{yu2011comparison}
Hao Yu, Tiantian Xie, Michael Hamilton, and Bogdan Wilamowski.
\newblock Comparison of different neural network architectures for digit image
  recognition.
\newblock In {\em 2011 4th International Conference on Human System
  Interactions, HSI 2011}, pages 98--103. IEEE, 2011.
\newblock \href {https://doi.org/10.1109/HSI.2011.5937350}
  {\path{doi:10.1109/HSI.2011.5937350}}.

\bibitem{cordoni2020comparison}
F~Cordoni.
\newblock A comparison of modern deep neural network architectures for energy
  spot price forecasting.
\newblock {\em Digital Finance}, 2:189--210, 2020.
\newblock \href {https://doi.org/10.1007/s42521-020-00022-2}
  {\path{doi:10.1007/s42521-020-00022-2}}.

\bibitem{szandala2021review}
Tomasz Szanda{\l}a.
\newblock Review and comparison of commonly used activation functions for deep
  neural networks.
\newblock In {\em Bio-inspired Neurocomputing}, pages 203--224. Springer, 2021.

\bibitem{nwankpa2018activation}
Chigozie Nwankpa, Winifred Ijomah, Anthony Gachagan, and Stephen Marshall.
\newblock Activation functions: Comparison of trends in practice and research
  for deep learning.
\newblock {\em arXiv preprint arXiv:1811.03378}, 2018.

\bibitem{urban2018neural}
Sebastian Urban.
\newblock {\em Neural network architectures and activation functions: A
  gaussian process approach}.
\newblock PhD thesis, Technische Universit{\"a}t M{\"u}nchen, 2018.

\bibitem{smith2017cyclical}
Leslie~N Smith.
\newblock Cyclical learning rates for training neural networks.
\newblock In {\em 2017 IEEE winter conference on applications of computer
  vision (WACV)}, pages 464--472. IEEE, 2017.
\newblock \href {https://doi.org/10.1109/WACV.2017.58}
  {\path{doi:10.1109/WACV.2017.58}}.

\bibitem{elsken2019neural}
Thomas Elsken, Jan~Hendrik Metzen, and Frank Hutter.
\newblock Neural architecture search: A survey.
\newblock {\em The Journal of Machine Learning Research}, 20(1):1997--2017,
  2019.

\bibitem{hutter2019automated}
Frank Hutter, Lars Kotthoff, and Joaquin Vanschoren.
\newblock {\em Automated machine learning: methods, systems, challenges}.
\newblock Springer Nature, 2019.
\newblock \href {https://doi.org/10.1007/978-3-030-05318-5}
  {\path{doi:10.1007/978-3-030-05318-5}}.

\bibitem{mcclean2018barren}
Jarrod~R McClean, Sergio Boixo, Vadim~N Smelyanskiy, Ryan Babbush, and Hartmut
  Neven.
\newblock Barren plateaus in quantum neural network training landscapes.
\newblock {\em Nature communications}, 9(1):1--6, 2018.
\newblock \href {https://doi.org/10.1038/s41467-018-07090-4}
  {\path{doi:10.1038/s41467-018-07090-4}}.

\bibitem{kiani2020learning}
Bobak~Toussi Kiani, Seth Lloyd, and Reevu Maity.
\newblock Learning unitaries by gradient descent.
\newblock {\em arXiv preprint arXiv:2001.11897}, 2020.

\bibitem{wiersema2020exploring}
Roeland Wiersema, Cunlu Zhou, Yvette de~Sereville, Juan~Felipe Carrasquilla,
  Yong~Baek Kim, and Henry Yuen.
\newblock Exploring entanglement and optimization within the hamiltonian
  variational ansatz.
\newblock {\em PRX Quantum}, 1(2):020319, 2020.
\newblock \href {https://doi.org/10.1103/PRXQuantum.1.020319}
  {\path{doi:10.1103/PRXQuantum.1.020319}}.

\bibitem{cerezo2020cost}
M~Cerezo, Akira Sone, Tyler Volkoff, Lukasz Cincio, and Patrick~J Coles.
\newblock Cost function dependent barren plateaus in shallow parametrized
  quantum circuits.
\newblock {\em Nature Communications}, 12(1):1--12, 2021.
\newblock \href {https://doi.org/10.1038/s41467-021-21728-w}
  {\path{doi:10.1038/s41467-021-21728-w}}.

\bibitem{wang2020noise}
Samson Wang, Enrico Fontana, Marco Cerezo, Kunal Sharma, Akira Sone, Lukasz
  Cincio, and Patrick~J Coles.
\newblock Noise-induced barren plateaus in variational quantum algorithms.
\newblock {\em Nature communications}, 12(1):1--11, 2021.
\newblock \href {https://doi.org/10.1038/s41467-021-27045-6}
  {\path{doi:10.1038/s41467-021-27045-6}}.

\bibitem{skolik2020layerwise}
Andrea Skolik, Jarrod~R McClean, Masoud Mohseni, Patrick van~der Smagt, and
  Martin Leib.
\newblock Layerwise learning for quantum neural networks.
\newblock {\em Quantum Machine Intelligence}, 3 (1):1--11, 2021.
\newblock \href {https://doi.org/10.1007/s42484-020-00036-4}
  {\path{doi:10.1007/s42484-020-00036-4}}.

\bibitem{marrero2020entanglement}
Carlos~Ortiz Marrero, M{\'a}ria Kieferov{\'a}, and Nathan Wiebe.
\newblock Entanglement-induced barren plateaus.
\newblock {\em PRX Quantum}, 2(4):040316, 2021.
\newblock \href {https://doi.org/10.1103/PRXQuantum.2.040316}
  {\path{doi:10.1103/PRXQuantum.2.040316}}.

\bibitem{sim2019expressibility}
Sukin Sim, Peter~D Johnson, and Al{\'a}n Aspuru-Guzik.
\newblock Expressibility and entangling capability of parameterized quantum
  circuits for hybrid quantum-classical algorithms.
\newblock {\em Advanced Quantum Technologies}, 2(12):1900070, 2019.
\newblock \href {https://doi.org/10.1002/qute.201900070}
  {\path{doi:10.1002/qute.201900070}}.

\bibitem{sim2021adaptive}
Sukin Sim, Jhonathan~Romero Fontalvo, J{\'e}r{\^o}me~F Gonthier, and
  Alexander~A Kunitsa.
\newblock Adaptive pruning-based optimization of parameterized quantum
  circuits.
\newblock {\em Quantum Science and Technology}, 2021.
\newblock \href {https://doi.org/10.1088/2058-9565/abe107}
  {\path{doi:10.1088/2058-9565/abe107}}.

\bibitem{liu2021layer}
Xiaoyuan Liu, Anthony Angone, Ruslan Shaydulin, Ilya Safro, Yuri Alexeev, and
  Lukasz Cincio.
\newblock Layer vqe: A variational approach for combinatorial optimization on
  noisy quantum computers.
\newblock {\em arXiv preprint arXiv:2102.05566}, 2021.
\newblock \href {https://doi.org/10.1109/TQE.2021.3140190}
  {\path{doi:10.1109/TQE.2021.3140190}}.

\bibitem{schuld2020effect}
Maria Schuld, Ryan Sweke, and Johannes~Jakob Meyer.
\newblock Effect of data encoding on the expressive power of variational
  quantum-machine-learning models.
\newblock {\em Physical Review A}, 103(3):032430, 2021.
\newblock \href {https://doi.org/10.1103/PhysRevA.103.032430}
  {\path{doi:10.1103/PhysRevA.103.032430}}.

\bibitem{cp_wiki}
Openai gym wiki, cartpole v0.
\newblock URL: \url{https://github.com/openai/gym/wiki/CartPole-v0}.

\bibitem{brockman2016openai}
Greg Brockman, Vicki Cheung, Ludwig Pettersson, Jonas Schneider, John Schulman,
  Jie Tang, and Wojciech Zaremba.
\newblock Openai gym.
\newblock {\em arXiv preprint arXiv:1606.01540}, 2016.

\bibitem{perez2020data}
Adri{\'a}n P{\'e}rez-Salinas, Alba Cervera-Lierta, Elies Gil-Fuster, and
  Jos{\'e}~I Latorre.
\newblock Data re-uploading for a universal quantum classifier.
\newblock {\em Quantum}, 4:226, 2020.
\newblock \href {https://doi.org/10.22331/q-2020-02-06-226}
  {\path{doi:10.22331/q-2020-02-06-226}}.

\bibitem{ota2021training}
Kei Ota, Devesh~K Jha, and Asako Kanezaki.
\newblock Training larger networks for deep reinforcement learning.
\newblock {\em arXiv preprint arXiv:2102.07920}, 2021.

\bibitem{quantum_agents_code}
Code used in this work https://github.com/askolik/quantum\_agents.
\newblock URL: \url{https://github.com/askolik/quantum_agents}.

\bibitem{sutton2018reinforcement}
Richard~S Sutton and Andrew~G Barto.
\newblock {\em Reinforcement learning: An introduction}.
\newblock MIT press, 2018.
\newblock \href {https://doi.org/10.1109/TNN.1998.712192}
  {\path{doi:10.1109/TNN.1998.712192}}.

\bibitem{sutton1999policy}
Richard~S Sutton, David~A McAllester, Satinder~P Singh, Yishay Mansour, et~al.
\newblock Policy gradient methods for reinforcement learning with function
  approximation.
\newblock In {\em NIPs}, volume~99, pages 1057--1063. Citeseer, 1999.

\bibitem{greensmith2004variance}
Evan Greensmith, Peter~L Bartlett, and Jonathan Baxter.
\newblock Variance reduction techniques for gradient estimates in reinforcement
  learning.
\newblock {\em Journal of Machine Learning Research}, 5(9), 2004.

\bibitem{melo2001convergence}
Francisco~S Melo.
\newblock Convergence of q-learning: A simple proof.
\newblock {\em Institute Of Systems and Robotics, Tech. Rep}, pages 1--4, 2001.

\bibitem{lin1992self}
Long-Ji Lin.
\newblock {\em Self-supervised Learning by Reinforcement and Artificial Neural
  Networks}.
\newblock PhD thesis, Carnegie Mellon University, School of Computer Science,
  1992.

\bibitem{melo2007q}
Francisco~S Melo and M~Isabel Ribeiro.
\newblock Q-learning with linear function approximation.
\newblock In {\em International Conference on Computational Learning Theory},
  pages 308--322. Springer, 2007.
\newblock \href {https://doi.org/10.1007/978-3-540-72927-3_23}
  {\path{doi:10.1007/978-3-540-72927-3_23}}.

\bibitem{kandala2017hardware}
Abhinav Kandala, Antonio Mezzacapo, Kristan Temme, Maika Takita, Markus Brink,
  Jerry~M Chow, and Jay~M Gambetta.
\newblock Hardware-efficient variational quantum eigensolver for small
  molecules and quantum magnets.
\newblock {\em Nature}, 549(7671):242--246, 2017.
\newblock \href {https://doi.org/10.1038/nature23879}
  {\path{doi:10.1038/nature23879}}.

\bibitem{liu2021rigorous}
Yunchao Liu, Srinivasan Arunachalam, and Kristan Temme.
\newblock A rigorous and robust quantum speed-up in supervised machine
  learning.
\newblock {\em Nature Physics}, pages 1--5, 2021.
\newblock \href {https://doi.org/10.1038/s41567-021-01287-z}
  {\path{doi:10.1038/s41567-021-01287-z}}.

\bibitem{dunjko2017exponential}
Vedran Dunjko, Yi-Kai Liu, Xingyao Wu, and Jacob~M Taylor.
\newblock Exponential improvements for quantum-accessible reinforcement
  learning.
\newblock {\em arXiv preprint arXiv:1710.11160}, 2017.

\bibitem{shor1999polynomial}
Peter~W Shor.
\newblock Polynomial-time algorithms for prime factorization and discrete
  logarithms on a quantum computer.
\newblock {\em SIAM review}, 41(2):303--332, 1999.
\newblock \href {https://doi.org/10.1137/S0036144598347011}
  {\path{doi:10.1137/S0036144598347011}}.

\bibitem{fl_wiki}
Openai gym wiki, frozen lake v0.
\newblock URL: \url{https://github.com/openai/gym/wiki/FrozenLake-v0}.

\bibitem{broughton2020tensorflow}
Michael Broughton, Guillaume Verdon, Trevor McCourt, Antonio~J Martinez,
  Jae~Hyeon Yoo, Sergei~V Isakov, Philip Massey, Murphy~Yuezhen Niu, Ramin
  Halavati, Evan Peters, et~al.
\newblock Tensorflow quantum: A software framework for quantum machine
  learning.
\newblock {\em arXiv preprint arXiv:2003.02989}, 2020.

\bibitem{cirq}
Cirq, https://quantumai.google/cirq.
\newblock URL: \url{https://quantumai.google/cirq}.

\bibitem{gym_leaderboard}
Openai gym leaderboard.
\newblock URL: \url{https://github.com/openai/gym/wiki/Leaderboard}.

\bibitem{liu2018differentiable}
Jin-Guo Liu and Lei Wang.
\newblock Differentiable learning of quantum circuit born machines.
\newblock {\em Physical Review A}, 98(6):062324, 2018.
\newblock \href {https://doi.org/10.1103/PhysRevA.98.062324}
  {\path{doi:10.1103/PhysRevA.98.062324}}.

\end{thebibliography}

\onecolumngrid
\appendix

\section{Visualization of a learned Q-function}
\label{sec:appendix_vis}

\begin{figure}[h]
	\includegraphics[width=\textwidth]{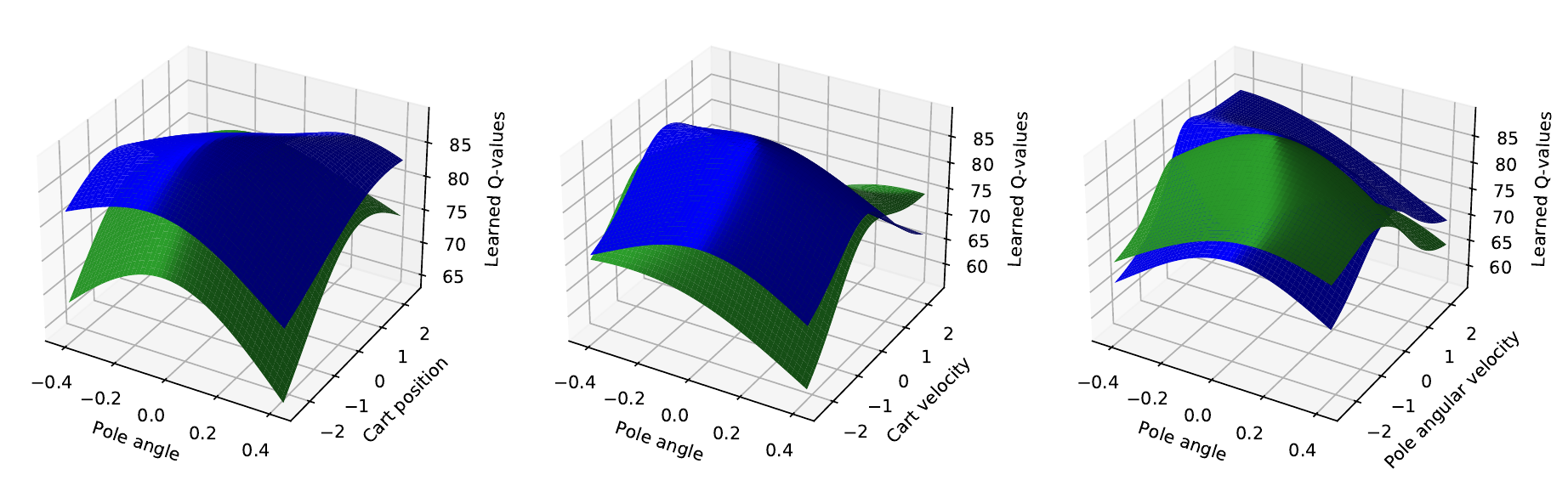}
  \caption{Visualization of the approximate Q-function learned by a quantum Q-learning agent solving Cart Pole. Due to the $4$ dimensions of the state space in Cart Pole, we represent the Q-values associated to the actions ``left'' (green) and ``right'' (blue) on $3$ subspaces of the state space by fixing unrepresented dimensions to $0$ in each plot. As opposed to the analogue values (i.e., unnormalized policy) learned by policy-gradient PQC agents in this environment \cite{jerbi2021variational}, the approximate Q-values appear nicely-behaved, likely due to the stronger constraints that Q-learning has on well-performing function approximations.}
  \label{fig:vis-Q-function}
\end{figure}

\section{Model hyperparameters}
\label{sec:appendix_hps}

In the following, we give a detailed list of the hyperparameters for each configuration in \cref{fig:fl_results}, \cref{fig:cp_results_dataenc}, \cref{fig:cp_results_readout}, \cref{fig:cp_nn_comparison} and \cref{fig:best_pqc_nn}. The hyperparameters that we searched over for each model were the following (see explanations of each hyperparameter in \cref{tab:hp_descriptions}):

\begin{itemize}
    \item \textit{Frozen Lake v0}: update model, update target model, $\eta$
    \item \textit{Cart Pole v0, quantum model}: batch size, update model, update target model, $\eta$, train $w_d$, train $w_o$, $\eta_{w_d}$, $\eta_{w_o}$
    \item \textit{Cart Pole v0, classical model}: number of units per layer, batch size, update model, update target model, $\eta$
\end{itemize}

\bgroup
\def\arraystretch{1.5}
\begin{table}
\begin{center}
\captionsetup{justification=raggedright}
\resizebox{0.8\textwidth}{!}{
\begin{tabular}{|l|l|}
\hline
                      & \textbf{Hyperparameter explanation}                                                   \\ \hline
qubits                & number of qubits in circuit                                                           \\ \hline
layers                & number of layers as defined in \cref{fig:pqc}                                         \\ \hline
$\gamma$              & discount factor for Q-learning                                                        \\ \hline
train $w_d$           & train weights on the model input as defined in \cref{sec:encoding}      \\ \hline
train $w_o$           & train weights on the model output as defined in \cref{sec:computing_qs}                                        \\ \hline
$\eta$          & model parameter learning rate                                                         \\ \hline
$\eta_{w_d}$        & input weight learning rate                                                            \\ \hline
$\eta_{w_o}$          & output weight learning rate                                                           \\ \hline
batch size    & number of samples shown to optimizer at each update                                   \\ \hline
$\epsilon_{\mathrm{init}}$                      & initial value for $\epsilon$-greedy policy                                               \\ \hline
$\epsilon_{\mathrm{dec}}$                       & decay of $\epsilon$ for $\epsilon$-greedy policy                                            \\ \hline
$\epsilon_{\mathrm{min}}$                 & minimal value of $\epsilon$ for $\epsilon$-greedy policy                                    \\ \hline
update model          & time steps after which model is updated                                         \\ \hline
update target model    & time steps after which model parameters are copied to target model              \\ \hline
size of replay memory &  size of memory for experience replay                                                  \\ \hline
data re-uploading        & use data re-uploading as defined in \cref{sec:encoding} \\ \hline
\end{tabular}
}
\caption{Description of hyperparameters considered in this work}
\label{tab:hp_descriptions} 
\end{center}
\end{table}
\egroup

\bgroup
\def\arraystretch{1.5}
\begin{table}
\begin{center}
\captionsetup{justification=raggedright}
\resizebox{0.8\textwidth}{!}{
\begin{tabular}{|l|c|c|c|c|}
\hline
                      & \multicolumn{1}{l|}{\textbf{Frozen Lake v0, \cref{fig:fl_results}}}  & \multicolumn{1}{l|}{\textbf{Cart Pole v0, optimal}}     & \multicolumn{1}{l|}{\textbf{Cart Pole v0, sub-optimal}}    \\ \hline
qubits                & 4  & 4  & 4    \\ \hline
layers                & 5, 10, 15 & 5 & 5           \\ \hline
$\gamma$                 & 0.8       & 0.99 & 0.99      \\ \hline
train $w_d$             & no    & yes, no & yes, no          \\ \hline
train $w_o$             & no    & yes, no & yes, no                   \\ \hline
$\eta$                   & 0.001   & 0.001 & 0.001   \\ \hline
$\eta_{w_d}$               &  --  & 0.001 & 0.001                                    \\ \hline
$\eta_{w_o}$               &  --  & 0.1 & 0.1                                     \\ \hline
batch size            & 11      & 16 & 16                              \\ \hline
$\epsilon_{\mathrm{init}}$          & 1  & 1 & 1                                   \\ \hline
$\epsilon_{\mathrm{dec}}$         & 0.99    & 0.99 & 0.99      \\ \hline
$\epsilon_{\mathrm{min}}$           & 0.01  & 0.01 & 0.01        \\ \hline
update model          & 5  & 1 & 10                    \\ \hline
update target model   & 10  & 1 & 30         \\ \hline
size of replay memory & 10000  & 10000 & 10000           \\ \hline
data re-uploading     & no & yes, no & yes, no    \\ \hline
\end{tabular}
}
\caption{Hyperparameter settings of PQCs in \cref{fig:fl_results}, \cref{fig:cp_results_dataenc} and \cref{fig:cp_results_readout}}
\label{tab:pqc_hps} 
\end{center}
\end{table}
\egroup

\bgroup
\def\arraystretch{1.5}
\begin{table}
\begin{center}
\captionsetup{justification=raggedright}
\resizebox{0.5\textwidth}{!}{
\begin{tabular}{|l|c|c|c|c|c|c|}
\hline

layers  & 5 & 10 & 15 & 20 & 25 & 30 \\ \hline
qubits                & 4  & 4  & 4   & 4  & 4  & 4    \\ \hline
$\gamma$                 & 0.99       & 0.99 & 0.99  & 0.99       & 0.99 & 0.99      \\ \hline
train $w_d$             & yes    & yes & yes  & yes    & yes & yes          \\ \hline
train $w_o$             & yes    & yes & yes & yes    & yes & yes                   \\ \hline
$\eta$                   & 0.001   & 0.001 & 0.001   & 0.001   & 0.001 & 0.001   \\ \hline
$\eta_{w_d}$               &  0.001  & 0.001 & 0.001    & 0.001   & 0.001 & 0.001                                    \\ \hline
$\eta_{w_o}$               &   0.1 & 0.1 & 0.1 & 0.1   & 0.1 & 0.1                                     \\ \hline
batch size            & 16     & 64 & 32    & 16      & 64 & 16                              \\ \hline
$\epsilon_{\mathrm{init}}$          & 1  & 1 & 1     & 1  & 1 & 1                                   \\ \hline
$\epsilon_{\mathrm{dec}}$         & 0.99    & 0.99 & 0.99    & 0.99    & 0.99 & 0.99      \\ \hline
$\epsilon_{\mathrm{min}}$           & 0.01  & 0.01 & 0.01    & 0.01  & 0.01 & 0.01        \\ \hline
update model          & 1  & 10 & 10  & 10  & 10 & 10                    \\ \hline
update target model   & 1  & 30 & 30  & 30  & 30 & 30         \\ \hline
size of replay memory & 10000  & 10000 & 10000  & 10000  & 10000 & 10000           \\ \hline
data re-uploading     & yes & yes & yes  & yes & yes & yes    \\ \hline
\end{tabular}
}
\caption{Hyperparameter settings of PQCs in \cref{fig:cp_nn_comparison} a)}
\label{tab:pqc_hps_nn_comp} 
\end{center}
\end{table}
\egroup

\bgroup
\def\arraystretch{1.5}
\begin{table}
\begin{center}
\captionsetup{justification=raggedright}
\resizebox{0.6\textwidth}{!}{
\begin{tabular}{|l|c|c|c|c|c|c|}
\hline

units in hidden layers  & (10, 10) & (15, 15) & (20, 20) & (24, 24) & (30, 30) & (64, 64) \\ \hline
$\gamma$                 & 0.99       & 0.99 & 0.99  & 0.99       & 0.99 & 0.99      \\ \hline
$\eta$                   & 0.001   & 0.001 & 0.001   & 0.001   & 0.001 & 0.001   \\ \hline
batch size            & 64     & 16 & 64    & 64      & 64 & 16                              \\ \hline
$\epsilon_{\mathrm{init}}$          & 1  & 1 & 1     & 1  & 1 & 1                                   \\ \hline
$\epsilon_{\mathrm{dec}}$         & 0.99    & 0.99 & 0.99    & 0.99    & 0.99 & 0.99      \\ \hline
$\epsilon_{\mathrm{min}}$           & 0.01  & 0.01 & 0.01    & 0.01  & 0.01 & 0.01        \\ \hline
update model          & 1  & 1 & 1  & 1  & 1 & 1                    \\ \hline
update target model   & 1  & 1 & 1  & 1  & 1 & 1         \\ \hline
size of replay memory & 10000  & 10000 & 10000  & 10000  & 10000 & 10000           \\ \hline
\end{tabular}
}
\caption{Hyperparameter settings of NNs in \cref{fig:cp_nn_comparison} b)}
\label{tab:pqc_hps_nn} 
\end{center}
\end{table}
\egroup

\end{document}